# Physics-Informed Neural Networks for Radial Consolidation of Combined Electroosmotic, Vacuum and Surcharge Preloading Considering Smear Effects

Dong Li[a], Yapeng Cao[b*], Shuai Huang[c], Yujun Cui[d], Haiping Fu[a], Lu Yang[e], He Wei[f]

[a] *PhD, Department of Civil, Environmental, and Infrastructure Engineering, George Mason University, Fairfax, VA 22030, USA*

[b] *PhD, State Key Laboratory of Cryospheric Science and Frozen Soil Engineering, Northwest Institute of Eco-Environment and Resources, Chinese Academy of Sciences, Lanzhou 730000, China; Laboratoire Navier/CERMES, École Nationale des Ponts et Chaussées, Institut Polytechnique de Paris, 77455 Marne-la-Vallée cedex 2, France (Corresponding author) email:* caoyapeng@lzb.ac.cn

[c] *PhD, State Key Laboratory of Tunnel Engineering, China University of Mining and Technology (Beijing), Beijing 100083, China*

[d] *Professor, Laboratoire Navier/CERMES, École Nationale des Ponts et Chaussées, Institut Polytechnique de Paris, 77455 Marne-la-Vallée cedex 2, France*

[e] *Associate Professor, College of Water Conservancy and Hydropower Engineering, Hohai University, Nanjing 210098, China*

[f] *School of Geosciences and Info-physics, Central South University, Changsha, 410083, China*

** Corresponding author Email address: caoyapeng@lzb.ac.cn*

*The number of words in the main text: 10788.*
*The number of figures and tables: 18 figures and 3 tables.*

**Abstract:** This study develops a dimensionless multi-domain physics-informed neural network (PINN) framework for electro-osmotic radial consolidation considering smear effects and combined vacuum–surcharge loading. Three PINN-based models are investigated: a standard soft-constrained PINN (Std-PINN), a modified gated PINN (Mod-PINN), and a modified gated PINN with hard-constraint boundary encoding (Mod-HC-PINN). The models are evaluated against FEM reference solutions under four loading cases, including constant vacuum, exponential vacuum, exponential vacuum with ramp surcharge, and exponential vacuum with cyclic haversine surcharge. The results indicate that gated architecture applied in Mod-PINN improves the resolution of steep pressure gradients near the cathode and smear-zone interface under constant vacuum loading. Under time-dependent loading, the soft-constrained Mod-PINN shows reduced accuracy because it must learn multiple competing objectives simultaneously. The Mod-HC-PINN mitigates this issue by embedding the cathode boundary and initial conditions into the output structure, thereby reducing the optimization burden and improving physical consistency. The Mod-HC-PINN achieves MAE values of 0.43, 0.41, and 0.27 kPa for the exponential vacuum, ramp surcharge, and cyclic surcharge cases, respectively. Sensitivity analyses further demonstrate that the proposed framework remains robust across practical ranges of network architecture, collocation density, and permeability contrast.

**Keywords:** Electro-osmotic consolidation; Smear effect; Electroosmosis-vacuum-surcharge preloading; Physics-informed neural network; Hard-constraint boundary encoding

# Notation

| Symbol | Description | Unit |
|---|---|---|
| $r$ | Radial coordinate | m |
| $r_w$ | Drain/cathode radius | m |
| $r_s$ | Smear-zone outer radius | m |
| $r_e$ | Influence/anode radius | m |
| $t$ | Time | d |
| $T$ | Dimensionless time | — |
| $T_{final}$ | Total simulation time | d |
| $R$ | Dimensionless radial coordinate | — |
| $R_w$ | Dimensionless cathode radius | — |
| $R_s$ | Dimensionless smear-zone interface radius | — |
| $R_e$ | Dimensionless outer influence radius | — |
| $u_1$ | Excess pore-water pressure in the smear zone | kPa |
| $u_2$ | Excess pore-water pressure in the undisturbed zone | kPa |
| $U_1$ | Dimensionless excess pore-water pressure in the smear zone | — |
| $U_2$ | Dimensionless excess pore-water pressure in the undisturbed zone | — |
| $u_r$ | Reference excess pore-water pressure used for normalization | kPa |
| $\gamma_w$ | Unit weight of water | $kN/m^3$ |
| $k_{r1}$ | Hydraulic permeability coefficient of the smear zone | m/d |
| $k_{r2}$ | Hydraulic permeability coefficient of the undisturbed zone | m/d |
| $k_{e1}$ | Electro-osmotic permeability coefficient of the smear zone | $m^2/(V \cdot d)$ |
| $k_{e2}$ | Electro-osmotic permeability coefficient of the undisturbed zone | $m^2/(V \cdot d)$ |
| $m_{v1}$ | Coefficient of volume compressibility of the smear zone | 1/kPa |
| $m_{v2}$ | Coefficient of volume compressibility of the undisturbed zone | 1/kPa |
| $c_{v1}$ | Coefficient of consolidation of the smear zone | $m^2/d$ |
| $c_{v2}$ | Coefficient of consolidation of the undisturbed zone | $m^2/d$ |
| $V(r)$ | Electric potential distribution | V |
| $V_{max}$ | Applied voltage at the anode | V |
| $p(t)$ | Time-dependent vacuum pressure | kPa |
| $p_v$ | Stabilized vacuum pressure | kPa |
| $P(T)$ | Dimensionless vacuum pressure function | — |
| $\alpha$ | Vacuum rate parameter | 1/d |
| $\beta$ | Dimensionless vacuum rate parameter | — |

| Symbol | Description | Unit |
|---|---|---|
| $q(t)$ | Time-dependent surcharge load | kPa |
| $q_u$ | Final or maximum surcharge load | kPa |
| $t_c$ | Ramp loading duration | d |
| $t_0$ | Period of haversine cyclic loading | d |
| $v_r$ | Radial pore-water flow velocity | m/d |
| $\theta_1$ | Trainable parameters of the smear-zone sub-network | — |
| $\theta_2$ | Trainable parameters of the undisturbed-zone sub-network | — |
| $\hat{U}_1$ | PINN-predicted dimensionless pore pressure in the smear zone | — |
| $\hat{U}_2$ | PINN-predicted dimensionless pore pressure in the undisturbed zone | — |
| $H^{(l)}$ | Hidden state of the l-th neural-network layer | — |
| $U, V$ | Encoder-branch outputs in the modified MLP | — |
| $L_{STD}$ | Total loss of the standard soft-constrained PINN | — |
| $L_{HC}$ | Total loss of the hard-constrained PINN | — |
| $L_{PDE,1}$ | PDE residual loss in the smear zone | — |
| $L_{PDE,2}$ | PDE residual loss in the undisturbed zone | — |
| $L_{IC}$ | Initial-condition loss term | — |
| $L_{BC,c}$ | Cathode boundary-condition loss term | — |
| $L_{BC,a}$ | Anode boundary-condition loss term | — |
| $L_C$ | Interface continuity loss term | — |
| $\omega_j$ | Weighting coefficient for the j-th loss component | — |
| $f_i$ | PDE residual in zone i | — |
| $N_i$ | Number of interior collocation points in zone i | — |
| $N_{IC}$ | Number of initial-condition collocation points | — |
| $N_{BC}$ | Number of boundary-condition collocation points | — |
| $N_C$ | Number of interface collocation points | — |
| $\nabla_\theta L$ | Gradient of a loss term with respect to trainable parameters | — |
| $cos(\nabla_\theta L_i, \nabla_\theta L_j)$ | Gradient cosine similarity between two loss terms | — |
| MAE | Mean absolute error | kPa |
| MRE | Mean relative error | % |

*Note: "—" indicates a dimensionless quantity or not applicable unit.*

# 1. Introduction

Soft clay soils, characterized by high water content, low hydraulic conductivity, and high compressibility, are widely encountered in coastal reclamation projects, dredged deposits, tailings impoundments, and infrastructure founded on marine and lacustrine sediments. Ground improvement is therefore often required to satisfy stability and serviceability requirements prior to construction. Within an axisymmetric unit-cell framework, vertical drainage systems combined with surcharge loading and/or vacuum preloading are widely adopted to accelerate consolidation (Wang and Vu, 2010; Wang et al., 2016; Zhang et al., 2025a; Zhang et al., 2026). Vacuum preloading induces negative pore pressure through a sealed drainage system, thereby increasing effective stress without increasing total stress. Consequently, it promotes pore-water dissipation and consolidation while reducing the risk of instability associated with large surcharge loads (Kjellman, 1952). However, conventional ground improvement methods relying primarily on hydraulic drainage and surcharge loading, such as prefabricated vertical drains (PVDs) combined with preloading, often require prolonged treatment periods because of the inherently low hydraulic conductivity of fine-grained soils (Zhang et al., 2024b). Moreover, installation of vertical drains disturbs and remolds the surrounding soil, leading to the formation of a smear zone with reduced permeability due to installation-induced disturbance and soil structure alteration (Indraratna and Redana, 1998; Zhu and Yin, 2004; Walker and Indraratna, 2007). To further enhance treatment efficiency, combined improvement techniques have been proposed, among which electro-osmosis integrated with vacuum preloading has attracted increasing attention as a promising approach for accelerating dewatering and improving the bearing capacity of soft clay foundations (Wan and Mitchell, 1976; Su and Wang, 2003; Hu et al., 2012; Zong et al., 2024; Zhang et al., 2024b).

Electro-osmosis was first introduced into geotechnical engineering by Casagrande (1949), and the corresponding theory of electro-osmotic consolidation was subsequently developed by Esrig (1968) through a one-dimensional consolidation model. Subsequent studies examined additional mechanisms such as surcharge loading and electrode reversal (Wan and Mitchell, 1976) and extended the governing framework to two-dimensional planar conditions (Su and Wang, 2003; Hu et al., 2012). Electro-osmotic ground improvement systems commonly employ vertically installed electrodes, with prefabricated vertical drains (PVDs) and electric vertical drains (EVDs) arranged in an equilateral triangular pattern. For these configurations, where anodes surround a cathode in a hexagonal layout, axisymmetric modeling provides a more appropriate representation of consolidation behavior than planar formulations (Glendinning et al., 2008; Wu and Hu, 2012). Wu and Hu (2013) developed an axisymmetric electro-osmotic consolidation model that accounts for the coupled effects of horizontal and vertical flow, where the smear effects were ignored in the model. This limitation was subsequently addressed by Liu et al. (2022), who presented an analytical solution for axisymmetric electro-osmotic consolidation under free-strain conditions, in which distinct soil properties were assigned to the smear zone and the surrounding undisturbed soil. More recently, combined electro-osmosis-vacuum-surcharge preloading has attracted growing research interest (Wang et al., 2019; Wang et al., 2020; Zong et al., 2024; Zhang et al., 2024b). Wang et al. (2020) proposed an analytical axisymmetric consolidation model for EVDs subjected to combined electro-osmosis, vacuum preloading, and surcharge loading. Their solution incorporates coupled vertical and horizontal flow, smear effects, well resistance, time-dependent loading, and variation of additional stress with depth. Wang et al. (2021) further proposed a nonlinear consolidation model for electro-osmosis-vacuum-surcharge preloading, accounting for the nonlinear variation of compressibility, hydraulic permeability, and electro-osmotic permeability with void ratio under the equal-strain assumption. Under free-strain conditions and considering smear effects, Zong et al. (2024) derived an axisymmetric radial consolidation solution for combined electro-osmotic, vacuum, and surcharge preloading under linear loading, and Zhang et al. (2024b) further extended the analytical framework to cyclic surcharge loading. However, deriving closed-form analytical solutions generally requires advanced integral transforms and eigenvalue analysis, which makes the formulation mathematically involved and computationally cumbersome. Numerical techniques are therefore commonly adopted to solve the governing PDEs, including finite difference (FD) methods (Rittirong and Shang, 2008; Zhou et al., 2013) and finite element (FE) methods (Lewis and Humpheson, 1973; Hu et al., 2012; Yuan and Hicks, 2013; Wu et al., 2017). These methods have been extended to account for coupled hydraulic-electro-osmotic flow, large deformation, and nonlinear variation of soil properties, thereby providing useful numerical solutions for electro-osmotic consolidation analysis. Nevertheless, conventional numerical methods still face several limitations, including susceptibility to mesh distortion under large deformation and convergence difficulties under complex boundary and initial conditions (Reddy, 2026).

Recent advances in scientific machine learning have introduced Physics-Informed Neural Networks (PINNs) as a mesh-free method for solving partial differential equations by embedding governing equations, boundary conditions, and initial conditions into the loss function of a neural network (Cuomo et al., 2022; Lawal et al., 2022). This approach was pioneered by Raissi et al. (2019) and was subsequently applied by Bekele (2021) to Terzaghi's consolidation. Bekele (2021) demonstrated that a PINN can accurately reproduce the one-dimensional consolidation solution (forward problem) and estimate the consolidation coefficient from measured data (inverse problem) by minimizing a combined physics-based and data-driven loss. Zhang et al. (2024a) developed a physics-informed data-driven approach to automatically recover Terzaghi's consolidation theory from measured data and obtain accurate solutions, while incorporating weak-form PDEs for noise robustness and Monte Carlo dropout for uncertainty estimation. Lan et al. (2024) enhanced PINNs by incorporating hard constraints (PINNs-H) to model excess pore-water pressure variation in nonlinear consolidation with continuous drainage boundaries. Zhang et al. (2025b) further investigated one-dimensional soil consolidation under multi-stage surcharge loading, introducing a temporal domain-decomposition strategy to improve solution stability when drainage conditions and loading stages evolve over time. Xie et al. (2025) introduced a PINN framework for nonlinear large-strain consolidation under a top-applied load with exponentially time-varying drainage boundaries, using double-logarithmic laws for soil compressibility and permeability and a scaling transformation to mitigate length- and time-scale conditioning issues. Li et al. (2026) further extended PINN-based consolidation analysis to one-dimensional unsaturated soils under long-term loading by developing a lagged backward-compatible PINN (LBC-PINN), which combines logarithmic time segmentation, lagged compatibility enforcement, and transfer learning to improve stability and accuracy across multi-scale time domains. Recent studies have demonstrated the applicability of PINNs to consolidation problems under various initial and boundary conditions; however, most existing studies have focused on classical Terzaghi one-dimensional consolidation, typically assuming homogeneous soil conditions and simplified drainage boundaries.

In contrast, electro-osmotic consolidation is governed by a different mechanism, in which pore-water flow is driven by both hydraulic gradients and externally applied electric fields. This introduces additional complexities into the governing equations, particularly when combined with time-dependent surcharge loading and vacuum preloading. Furthermore, field applications commonly involve radially heterogeneous ground conditions, such as installation-induced smear zones around vertical drains or electrodes, where contrasts in hydraulic and electro-osmotic permeability require explicit enforcement of excess pore-water pressure and radial flux continuity at soil-interface boundaries. These challenges are not fully addressed in existing PINN frameworks and therefore motivate the development of a tailored PINN approach capable of handling coupled electro-osmotic effects, multi-domain geometries with interface conditions, and time-varying loading histories.

Accordingly, this study develops a dimensionless multi-domain physics-informed neural network framework for radial consolidation of combined electroosmosis-vacuum-surcharge preloading considering smear effects. Three PINN models are investigated: (1) a standard fully connected PINN with soft constraint enforcement (Std-PINN), (2) a modified gated PINN with soft constraint enforcement (Mod-PINN), and (3) a modified gated PINN with hard-constraint boundary encoding (Mod-HC-PINN). Four loading cases of increasing complexity are considered: (C1) electro-osmotic consolidation under constant vacuum loading; (C2) electro-osmotic consolidation under exponential vacuum loading; (C3) electro-osmotic consolidation under exponential vacuum and ramp surcharge loading; and (C4) electro-osmotic consolidation under exponential vacuum and cyclic haversine surcharge loading. The predictive performance of each model is evaluated against reference FEM solutions. Additional sensitivity analyses are conducted to assess the robustness of the proposed framework with respect to network architecture, collocation density, and hydraulic/electroosmotic permeability contrasts.

# 2. Problem Description

## 2.1 Governing equations and loading functions

An axisymmetric unit-cell is considered for a vertical electric drain (EVD) installed in saturated soft clay as shown in Fig. 1. The vertical EVD with radius $r_w$ is installed at the center of the influenced zone. A disturbed smear zone created during installation extends to radius $r_s$, and the outer influence boundary is located at $r_e$. The radial coordinate satisfies $r \in [r_w, r_e]$. In this study, the basic model assumptions are presented below: 1) the soil specimen is fully saturated; 2) the distribution of voltage, V (r), is independent of time and assumed to follow the law of conservation of electrical charge in the radial direction; 3) Darcy's law and Ohm's law are considered for the flow of water in the soil; 4) water flow is assumed to be one-dimensional in the radial direction within the horizontal plane; 5) soil parameters remain constant during consolidation; and 6) thermal and electrochemical effects are ignored.

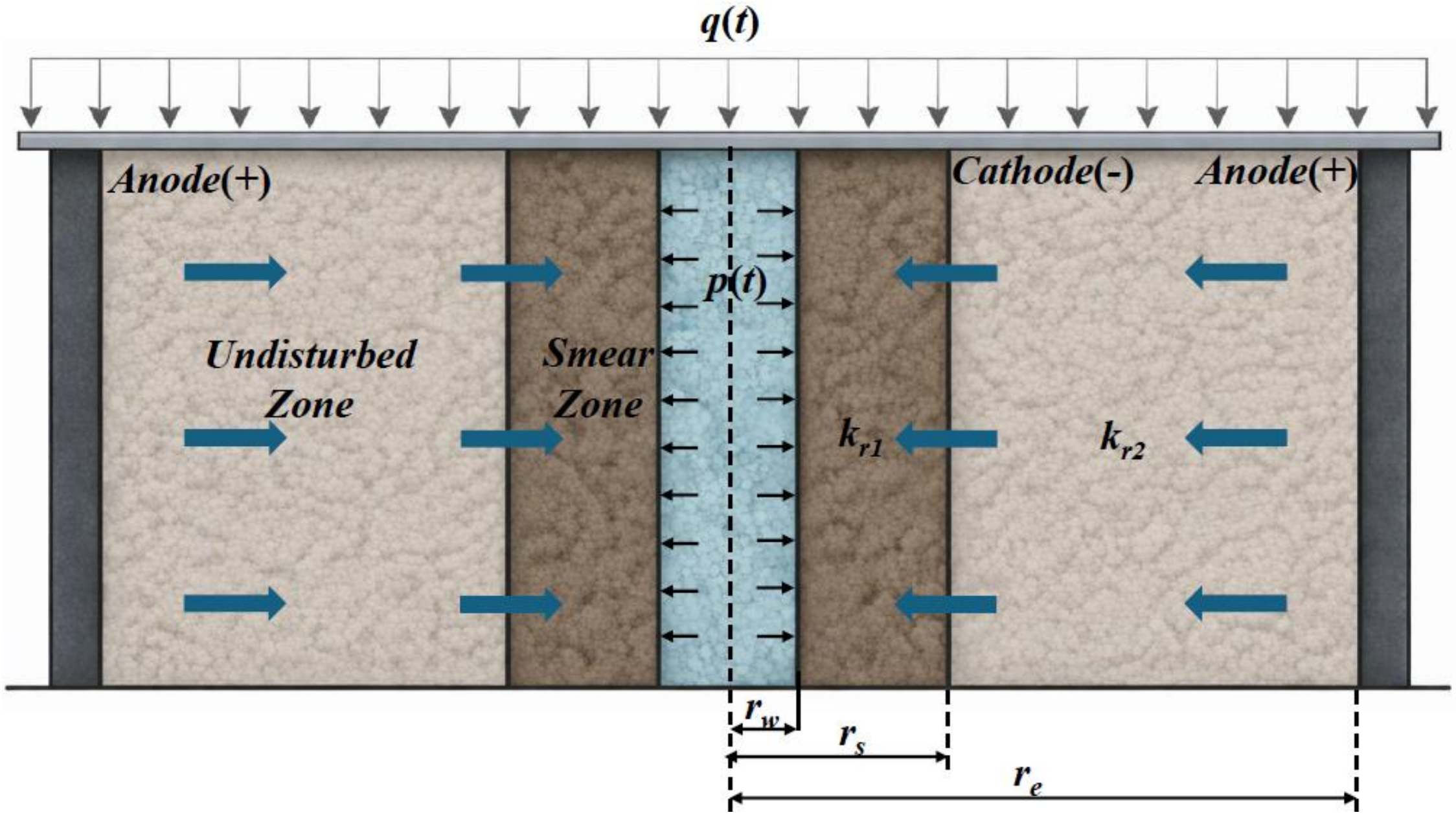

**Fig. 1.** Schematic illustration of radial consolidation under combined electro-osmotic, vacuum, and surcharge preloading.

To represent smear effects, the soil domain is decomposed into two concentric zones. 1) Zone 1 (inside the smear zone): $r \in (r_w, r_s]$ with excess pore-water pressure $u_1(r,t)$; 2) Zone 2 (undisturbed zone): $r \in [r_s, r_e)$ with excess pore-water pressure $u_2(r,t)$. For each zone $i = 1,2$, the axisymmetric radial electro-osmotic consolidation equation is written as:

$$\frac{k_{r1}}{\gamma_w m_{v1}}\left(\frac{1}{r}\frac{\partial u_1}{\partial r} + \frac{\partial^2 u_1}{\partial r^2}\right) + \frac{k_{e1}}{m_{v1}}\left(\frac{1}{r}\frac{\partial V}{\partial r} + \frac{\partial^2 V}{\partial r^2}\right) = \frac{\partial u_1}{\partial t} - \frac{dq(t)}{dt}, r \in (r_w, r_s] \tag{1}$$

$$\frac{k_{r2}}{\gamma_w m_{v2}}\left(\frac{1}{r}\frac{\partial u_2}{\partial r} + \frac{\partial^2 u_2}{\partial r^2}\right) + \frac{k_{e2}}{m_{v2}}\left(\frac{1}{r}\frac{\partial V}{\partial r} + \frac{\partial^2 V}{\partial r^2}\right) = \frac{\partial u_2}{\partial t} - \frac{dq(t)}{dt}, r \in [r_s, r_e) \tag{2}$$

where $k_{r1}$ and $k_{r2}$ denote the hydraulic permeability coefficients of the smear zone and undisturbed zone, respectively; $k_{e1}$ and $k_{e2}$ are the electroosmotic permeability coefficients of the smear zone and undisturbed zone, respectively. $u_1$ and $u_2$ represent the excess pore-water pressures in the smear zone and undisturbed zone, respectively. $m_{v1}$ and $m_{v2}$ are the coefficients of volume compressibility of the smear zone and undisturbed zone. The surcharge load history $q(t)$ appears through its time derivative $\mathrm{d}q/\mathrm{d}t$, representing the time-dependent surcharge loading in the governing equation.

The distribution of voltage function between anode and cathode under axisymmetric condition can be expressed as follows (Wu and Hu 2013):

$$V(r) = \frac{V_{max}}{ln(r_e/r_w)} ln\frac{r}{r_w} \tag{3}$$

In coupled electro-osmosis-vacuum-surcharge improvement, vacuum pressure is applied through the drain system and surcharge is applied at the ground surface. The vacuum pressure can be treated as either constant or time-dependent depending on the loading condition considered (Wang et al., 2021; Tian et al., 2022; Zhang et al., 2024a). For the time-dependent case, the vacuum pressure $p(t)$ can be approximated according to Tian et al. (2022) as follows:

$$p(t) = -p_v(1 - e^{-\alpha t}) \tag{4}$$

where $p_v$ is the applied vacuum pressure; $\alpha$ is a time effect parameter reflecting the rate of vacuum preloading.

The surcharge load history $q(t)$ can be constant, ramp surcharge loading and haversine cyclic loading (Liu et al., 2022; Zhang et al., 2024b; Zong et al., 2024). For the ramp loading case shown in Fig. 2(a), the applied surcharge load can be expressed as:

$$q(t) = \begin{cases} \frac{q_u}{t_c} * t, 0 \leq t \leq t_c \\ q_u, t_c \leq t \end{cases} \tag{5}$$

where $q_u$ is the final surcharge load, and $t$, $t_c$ denote time and the duration of single-stage surcharge application, respectively.

As shown in Fig. 2(b), the haversine cyclic loading may be written as:

$$q(t) = q_u \sin^2\left(\frac{\pi t}{t_0}\right) \tag{6}$$

where $q_u$ is the maximum cyclic surcharge load, and $t$, $t_0$ denote time and the period of one loading cycle, respectively.

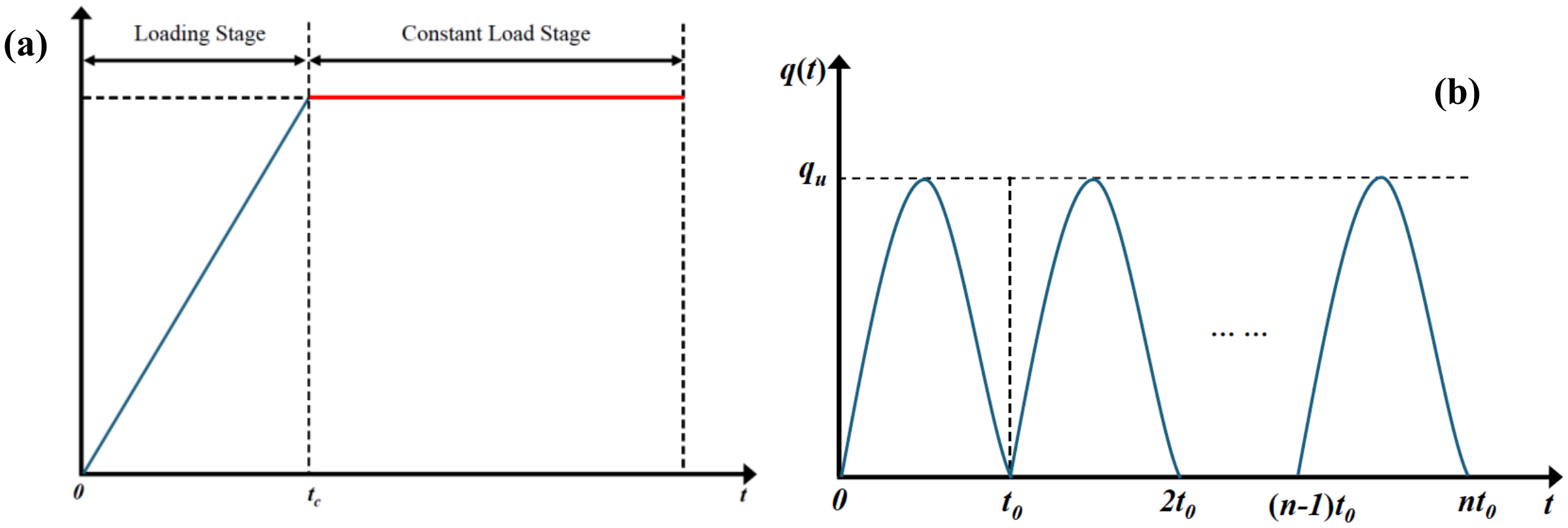

**Fig. 2.** Time-dependent surcharge loading patterns: (a) ramp-and-hold loading (Case C3); (b) cyclic loading (Case C4).

## 2.2 Initial, boundary, and interface conditions

The initial condition can be expressed as follows:

$$u_1(r,0) = u_2(r,0) = 0 \tag{7}$$

At the cathode boundary $r = r_w$, the vertical electric drain is assumed to be fully permeable, and the pore-water pressure equals the applied vacuum pressure $p(t)$; the electric potential is set as the reference potential (zero):

$$\text{Cathode } (r = r_w) \qquad u_1(r_w,t) = p(t), V(r_w,t) = 0 \tag{8}$$

At the anode boundary $r = r_e$, the total radial flux associated with hydraulic and electro-osmotic flow is assumed to be zero:

$$\text{Anode } (r = r_e) \qquad \frac{k_{r2}}{\gamma_w}\frac{\partial u_2(r_e,t)}{\partial r} + k_{e2}\frac{\partial V(r_e,t)}{\partial r} = 0 \tag{9}$$

At the smear-undisturbed zone interface $r = r_s$, continuity of excess pore-water pressure and continuity of total radial flux are enforced:

$$u_1(r_s,t) = u_2(r_s,t) \tag{10}$$

$$\frac{k_{r1}}{\gamma_w}\frac{\partial u_1(r_s,t)}{\partial r} + k_{e1}\frac{\partial V(r_s,t)}{\partial r} = \frac{k_{r2}}{\gamma_w}\frac{\partial u_2(r_s,t)}{\partial r} + k_{e2}\frac{\partial V(r_s,t)}{\partial r} \tag{11}$$

# 3. Physics-Informed Neural Network Method for Electro-Osmotic Radial Consolidation

This section presents the multi-domain physics-informed neural network (PINN) framework developed for axisymmetric electro-osmotic radial consolidation with smear effects. A PINN uses a differentiable neural network to approximate an unknown field, training it by minimizing PDE, initial condition, and boundary condition residuals at collocation points rather than requiring dense labeled data (Raissi et al., 2019; Karniadakis et al., 2021; Cuomo et al., 2022). Unlike mesh-based methods such as FEM and FDM, it evaluates the solution at arbitrary space-time coordinates and computes derivatives via automatic differentiation, yielding a mesh-free, differentiable representation of the consolidation process. Schematic diagrams of the three PINN frameworks are shown in Fig. 3.

Since soil properties differ between the smear and undisturbed zones, a single network cannot capture the derivative discontinuity at shared interface. To address this, a multi-domain PINN strategy is adopted, in which two separate networks are assigned to the respective zones and coupled through interface continuity constraints. The remainder of this section describes, in order, dimensionless formulation (Section 3.1), the domain decomposition and network parameterization (Section 3.2), the network architecture (Section 3.3), the hard-constraint boundary encoding (Section 3.4), the composite loss function (Section 3.5), and the collocation and sampling strategy (Section 3.6).

## 3.1 Dimensionless formulation

To enhance numerical conditioning and normalize all inputs and outputs to comparable scales, the governing equations in Section 2 are reformulated in dimensionless form before being embedded in the PINN loss function.

The following nondimensional variables are introduced:

$$R = \frac{r}{r_e}, \quad T = \frac{t}{t_f}, \quad U_i = \frac{u_i}{u_r}, \quad Q = \frac{q}{u_r}, \quad P = \frac{p}{u_r}, \hat{V} = \frac{V}{V_{Max}} \tag{12}$$

where $r_e$ is the outer influence radius, $t_f$ is the total simulation time, and $u_r$ is a reference excess pore-water pressure, mapping the computational domain to $R \in [R_w, 1]$ and $T \in [0,1]$.

Substituting Eq. (12) into the governing PDEs (Eq. (1-2)) yields the dimensionless consolidation equations:

Zone 1 (smear zone), $R \in (R_w, R_s]$**:**

$$C_{r1}\left(\frac{1}{R}\frac{\partial U_1}{\partial R} + \frac{\partial^2 U_1}{\partial R^2}\right) + C_{e1}\left(\frac{1}{R}\frac{d\tilde{V}}{dR} + \frac{d^2\tilde{V}}{dR^2}\right) = \frac{\partial U_1}{\partial T} - \frac{dQ}{dT} \tag{13}$$

Zone 2 (undisturbed zone), $R \in [R_s, 1)$:

$$C_{r2}\left(\frac{1}{R}\frac{\partial U_2}{\partial R} + \frac{\partial^2 U_2}{\partial R^2}\right) + C_{e2}\left(\frac{1}{R}\frac{d\tilde{V}}{dR} + \frac{d^2\tilde{V}}{dR^2}\right) = \frac{\partial U_2}{\partial T} - \frac{dQ}{dT} \tag{14}$$

in which four dimensionless consolidation coefficients are defined:

$$C_{ri} = \frac{k_{ri}\, t_f}{\gamma_w\, r_e^2\, m_{vi}}, \qquad C_{ei} = \frac{k_{ei}\, V_{max}\, t_f}{r_e^2\, m_{vi}\, u_r}, \qquad i = 1,\, 2 \tag{15}$$

The dimensionless initial, boundary, and interface conditions correspond to Eq. (7)- Eq. (11) become:

Initial condition:

$$U_1(R,\, 0) = U_2(R,\, 0) = 0 \tag{16}$$

Boundary conditions:

$$U_1(R_w,\, T) = P(T) \tag{17}$$

$$C_{r2}\frac{\partial U_2(1,\, T)}{\partial R} + C_{e2}\frac{d\tilde{V}(1)}{dR} = 0 \tag{18}$$

Interface continuity

$$U_1(R_s,\, T) = U_2(R_s,\, T) \tag{19}$$

$$C_{r1}\frac{\partial U_1}{\partial R}|_{R_s} + C_{e1}\frac{d\tilde{V}}{dR}|_{R_s} = C_{r2}\frac{\partial U_2}{\partial R}|_{R_s} + C_{e2}\frac{d\tilde{V}}{dR}|_{R_s} \tag{20}$$

To enable a unified hard-constraint formulation across cases C2-C4 (see Section 3.4), the cathode boundary condition adopts the exponential vacuum form for all three cases (cases C2-C4):

$$P(T) = -(p_v/u_r)\left(1 - e^{-\beta T}\right). \tag{21}$$

where $\beta$ is a dimensionless ramp-rate parameter, defined as $\beta = \alpha t_f$. The nondimensionalization maps the radial coordinate from $r \in [0.2,\ 1.8]$ m to $R \in [R_w,\ 1]$ and time from $t \in [0,\ 1000]$ days to $T \in [0,\ 1]$, eliminating the large disparity between physical length and time scales that would otherwise bias the neural-network gradients and cause training instability. The network output $U_i = u_i/u_r$ is also of order unity, promoting numerically stable training without output-layer rescaling.

## 3.2 Domain decomposition and network parameterization

Let $\widehat{N}_1(R, T;\, \boldsymbol{\theta}_1)$ and $\widehat{N}_2(R, T;\, \boldsymbol{\theta}_2)$ denote the network approximations to the dimensionless excess pore-water pressure in the smear zone ($R \in [R_w, R_s]$) and the undisturbed zone ($R \in [R_s, 1]$), respectively. Here $\boldsymbol{\theta}_1$ and $\boldsymbol{\theta}_2$ collect all trainable parameters (weights and biases) of each sub-network. During inference, the predicted dimensionless excess pore-water pressure at any point $(R, T)$ is obtained by the piecewise evaluation:

$$\widehat{U}(R, T) = \begin{cases} \widehat{N}_1(R, T; \boldsymbol{\theta}_1), & R_w \leq R \leq R_s \\ \widehat{N}_2(R, T; \boldsymbol{\theta}_2), & R_s < R \leq 1 \end{cases} \tag{22}$$

Each sub-network only needs to learn the solution within its own sub-domain, where the PDE coefficients are constant, reducing the complexity of the function to be approximated. For all required PDE residuals (Eq.13 and Eq.14), boundary condition (Eq.18), and interface continuity (Eq.20), the partial derivatives $\partial\widehat{N}_i/\partial R$, $\partial^2\widehat{N}_i/\partial R^2$, and $\partial\widehat{N}_i/\partial T$ are computed exactly using automatic differentiation (AD).

## 3.3 Network architectures: Std-MLP and Mod-MLP

Two network architectures are investigated in this work: a standard fully connected multilayer perceptron (Std-MLP) serving as the baseline and a Modified MLP (Mod-MLP) with multiplicative gating designed to address known training pathologies, as shown in Fig. 3. Both architectures use the same nominal depth, width, input dimension, and output dimension, allowing a controlled comparison of the effect of the internal connectivity pattern.

### 3.3.1 Standard fully connected MLP (Std-MLP)

The baseline architecture follows the original PINN formulation of Raissi et al. (2019) and employs a standard feedforward multilayer perceptron (MLP). In the Std-MLP, the input vector $\mathbf{x} = [R, T]^\top \in \mathbb{R}^2$ is first concatenated and then propagated through $L$ hidden layers according to:

$$\boldsymbol{h}^{(l)} = \sigma\left(\boldsymbol{W}^{(l)}\boldsymbol{h}^{(l-1)} + \boldsymbol{b}^{(l)}\right), \quad l = 1, 2, \ldots, L \tag{23}$$

where $\mathbf{h}^{(0)} = \mathbf{x}$ is the input; $\mathbf{W}^{(l)} \in \mathbb{R}^{n_l \times n_{l-1}}$ and $\mathbf{b}^{(l)} \in \mathbb{R}^{n_l}$ are the learnable weight matrix and bias vector of the $l$-th layer; $n_l$ is the layer width; and $\sigma(\cdot)$ denotes the elementwise activation function. The final hidden state is projected to a scalar output by a linear layer without activation:

$$\widehat{N} = \boldsymbol{w}^{(L+1)\top}\boldsymbol{h}^{(L)} + b^{(L+1)} \tag{24}$$

### 3.3.2 Modified MLP (Mod-MLP)

Standard fully-connected MLPs are known to suffer from two interrelated training pathologies when applied to PINNs for stiff or multi-scale PDEs. Standard MLPs exhibit a spectral bias toward low-frequency components, which limits their ability to resolve sharp pore-pressure gradients near the drain and across the smear-zone interface. Additionally, the composite PINN loss terms converge at different rates, leading to multi-objective optimization difficulties and training stagnation (Wang et al., 2023, 2024).

To address these challenges, this study also adopts a modified MLP architecture proposed by Wang et al. (2024). The key innovation is the introduction of two encoder branches ($\boldsymbol{U}$ and $\boldsymbol{V}$) that independently transform the raw input into the hidden-layer dimension, followed by a multiplicative gating mechanism at every hidden layer that replaces the standard sequential composition. The concatenated input $\boldsymbol{x} = [R,\, T]^{\top} \in \mathbb{R}^2$ is independently projected through two parallel encoder branches, each consisting of a linear transformation followed by a tanh activation:

$$\mathbf{U} = \tanh(\mathbf{W}_{\mathrm{U}}\, \mathbf{x} + \mathbf{b}_{\mathrm{U}}) \in \mathbb{R}^n \tag{25}$$

$$\mathbf{V} = \tanh(\mathbf{W}_{\mathrm{V}}\, \mathbf{x} + \mathbf{b}_{\mathrm{V}}) \in \mathbb{R}^n \tag{26}$$

where $\mathbf{W}_{\mathrm{U}}, \mathbf{W}_{\mathrm{V}} \in \mathbb{R}^{n\times 2}$ and $\mathbf{b}_{\mathrm{U}}, \mathbf{b}_{\mathrm{V}} \in \mathbb{R}^n$ are learnable parameters. The two encoder branches are initialized independently and are free to learn distinct nonlinear projections of the input space.

The hidden state is initialized as $\mathbf{H}^{(0)} = \mathbf{U}$. At each subsequent hidden layer k = 1, 2, …, L, a standard affine transformation followed by tanh activation produces an intermediate gate signal $\mathrm{Z}^{(\mathrm{k})}$:

$$\boldsymbol{Z}^{(\boldsymbol{k})} = \tanh\left(\boldsymbol{W}^{(k)}\, \boldsymbol{H}^{(k-1)} + \boldsymbol{b}^{(k)}\right) \in \mathbb{R}^n \tag{27}$$

where $\mathbf{W}^{(k)} \in \mathbb{R}^{n\times n}$ and $\mathbf{b}^{(k)} \in \mathbb{R}^n$ are the learnable weight matrix and bias vector of the $k$-th gated layer. This gate signal is then combined with the two encoder outputs through an element-wise gated blending operation:

$$\mathbf{H}^{(\mathbf{k})} = \left(\mathbf{1} - \mathbf{Z}^{(\mathbf{k})}\right) \odot \mathbf{U} + \mathbf{Z}^{(\mathbf{k})} \odot \mathbf{V} \tag{28}$$

where $\odot$ denotes the Hadamard (elementwise) product and $\mathbf{1}$ is a vector of ones. Because $\mathbf{Z}^{(k)}$ is generated by a tanh activation, its components lie in (-1,1). Each element of $\mathbf{H}^{(\mathbf{k})}$ represents an adaptive gated blending of the corresponding elements of $\boldsymbol{U}$ and $\boldsymbol{V}$. The gated blending coefficients vary across neurons, layers, and different input points $(R, T)$, enabling the network to adaptively select from the two encoder representations depending on the local characteristics of the solution. The final hidden state $\mathbf{H}^{(n)}$ is mapped to a scalar output by a linear layer without activation:

$$\widehat{\boldsymbol{U}} = \boldsymbol{w}_{\text{out}}^{\top} \boldsymbol{H}^{(\text{n})} + \boldsymbol{b}_{\text{out}} \tag{29}$$

## 3.4 Hard-constraint boundary encoding

In the standard PINN formulation, the initial condition (Eq. 16) and the cathode boundary condition (Eq. 17) are enforced as soft penalty terms within the composite loss function. When the cathode boundary is time-dependent, the optimizer must simultaneously learn the functional form P(T) along the cathode and propagate its effect into the interior solution while maintaining the zero initial condition and satisfying the PDE residual. This multi-objective task competes for finite network capacity, and numerical experiments in Section 5.2 show that standard soft-constraint formulations cannot allocate sufficient capacity to all objectives simultaneously, leading to systematic under-resolution of the PDE residual in the interior. The hard-constraint boundary encoding described below eliminates the boundary-condition and initial-condition loss terms from the loss by embedding their satisfaction directly into the network output structure, redirecting the full network capacity to the PDE residual, anode boundary, and interface continuity.

To resolve this capacity limitation, a hard-constraint output transformation is introduced that embeds the cathode Dirichlet boundary condition and the initial condition directly into the network output structure (Lu et al., 2021; Lagaris et al., 1998). Rather than penalizing violations of these conditions in the loss function, the predicted excess pore-water pressure is constructed so that the boundary and initial values are satisfied identically for any set of network parameters. Specifically, let $\widehat{N}_1$ and $\widehat{N}_2$ denote the raw outputs of the two sub-networks. The physically admissible predictions $\widehat{U}_1$ and $\widehat{U}_2$ are then constructed as follows:

Zone 1 (smear zone), $R \in [R_w, R_s]$**:**

$$\widehat{U}_1(R,T) = P(T) + (R - R_w) \cdot T \cdot \widehat{N}_1(R,T;\boldsymbol{\theta}_1) \tag{30}$$

Zone 2 (undisturbed zone), $R \in [R_s, 1)$:

$$\widehat{U}_2(R,T) = T \cdot \widehat{N}_2(R,T;\boldsymbol{\theta}_2) \tag{31}$$

At the cathode boundary R = $R_w$, substituting into Eq. (30)

$$\widehat{U}_1(R_w,T) = P(T) \tag{32}$$

At initial time T = 0, substituting into Eqs. (30) and (31):

$$\widehat{U}_1(R,0) = \widehat{U}_2(R,0) = 0 \tag{33}$$

The anode boundary condition (Eq. 18) and interface continuity (Eq. (19) and (20)) remain as soft penalty terms, as they involve derivative conditions or inter-network coupling that cannot be embedded through simple distance functions. The hard constraint transform was not applied to Case C1 because its constant nonzero vacuum pressure (e.g., $p(t) = p_v$) is incompatible with the zero initial condition.

## 3.5 Composite loss function formulation

The trainable parameters of the two sub-networks, $(\boldsymbol{\theta_1}, \boldsymbol{\theta_2})$, are identified by minimizing a composite loss function assembled from the governing-equation residuals, the initial condition, the boundary conditions, and the interface continuity constraints. Because three PINN models are evaluated in this study, two related objective functions are adopted. The Std-PINN and Mod-PINN use a soft-constraint loss, whereas the Mod-HC-PINN uses reduced loss terms in which the initial condition and cathode Dirichlet boundary condition are satisfied exactly by construction.

For the soft-constraint models (Std-PINN and Mod-PINN), the objective function is written as

$$\mathcal{L}_{Soft}(\boldsymbol{\theta}) = \omega_{PDE,1}\mathcal{L}_{PDE,1} + \omega_{PDE,2}\mathcal{L}_{PDE,2} + \omega_{IC}\,\mathcal{L}_{IC} + \omega_{BC,c}\mathcal{L}_{BC,c} + \omega_{BC,a}\,\mathcal{L}_{BC,a} + \omega_c\,\mathcal{L}_c \tag{34}$$

where $\omega(\cdot)$ are weighting coefficients, $\mathcal{L}_{PDE,1}$ and $\mathcal{L}_{PDE,2}$ are the PDE residual losses in the smear and undisturbed zones, respectively, $\mathcal{L}_{IC}$ is the initial-condition loss, $\mathcal{L}_{BC,c}$, $\mathcal{L}_{BC,a}$ are the cathode and anode boundary condition loss term, respectively, and $\mathcal{L}_C$ is the interface continuity loss.

For the hard-constraint model (Mod-HC-PINN), Eqs. (32) and (33) embed the initial condition and cathode Dirichlet boundary condition directly into the network output. Accordingly, the corresponding loss terms are removed from the optimization objective, which becomes:

$$\mathcal{L}_{HC}(\boldsymbol{\theta}) = \mathcal{L}_{PDE,1} + \mathcal{L}_{PDE,2} + \omega_{BC,a}\,\mathcal{L}_{BC,a} + \omega_c\,\mathcal{L}_c \tag{35}$$

where $\mathcal{L}_{BC,a}$is the anode boundary loss term.

For zone $i$, let $\hat{f}_i(R_j, T_j)$ denote the residual obtained by substituting the network prediction $\widehat{U}_i$ and its automatic-differentiation derivatives into the corresponding dimensionless governing equation, namely, Eq. (13) for Zone 1 and Eq. (14) for Zone 2. The zone-wise PDE residual losses are then defined as:

$$\mathcal{L}_{PDE,i} = \frac{1}{N_{PDE,i}} \sum_{j=1}^{N_{PDE,i}} [\hat{f}_i(R_j, T_j)]^2 \tag{36}$$

where $N_{PDE,i}$ is the number of interior collocation points sampled in zone $i$. For loading cases involving surcharge, the residual $\hat{f}_i$ includes the dimensionless source term associated with $dQ/dT$.

The initial-condition loss is evaluated at $N_{IC}$ collocation points distributed along $T = 0$:

$$\mathcal{L}_{IC} = \frac{1}{N_{IC}} \sum_{j=1}^{N_{IC}} [\,\widehat{U}(R_j, 0)]^2 \tag{37}$$

The cathode boundary loss is imposed at the drain boundary and enforces the prescribed vacuum pressure history:

$$\mathcal{L}_{BC,c} = \frac{1}{N_{BC,c}} \sum_{j=1}^{N_{BC,c}} \left[\widehat{U}_1(R_W, T_j) - P\,(T_j)\right]^2 \tag{38}$$

where $R_W$ is the dimensionless cathode radius and $P\,(T_j)$ is the dimensionless vacuum-loading function defined in Eq. (21).

Similarly, the anode boundary loss can be written as:

$$\mathcal{L}_{BC,a} = \frac{1}{N_{BC,a}} \sum_{j=1}^{N_{BC,a}} \left[ \frac{\partial \widehat{U}_2}{\partial R} |_{(1,T_j)} + \frac{C_{e2}}{C_{r2}} \frac{d\tilde{V}}{dR} |_{R=1} \right]^2 \tag{39}$$

The interface continuity loss ($\mathcal{L}_c$) can be written as:

$$\mathcal{L}_c = \frac{1}{N_s} \sum_{j=1}^{N_s} \left( \left[ \widehat{U}_1(R_s, T_j) - \widehat{U}_2(R_s, T_j) \right]^2 + \left[ \Delta \hat{q}_s(T_j) \right]^2 \right) \tag{40}$$

where $\Delta \hat{q}_s(T_j)$ denote as:

$$\Delta \hat{q}_s(T_j) = \sum_{j=1}^{N_s} \left[ C_{r1} \frac{\partial \widehat{U}_1}{\partial R} - C_{r2} \frac{\partial \widehat{U}_2}{\partial R} + (C_{e1} - C_{e2}) \frac{d\tilde{V}}{dR} \right]^2_{(R_s, T_j)} \tag{41}$$

To diagnose the interaction among different loss components during training, the cosine similarity between their parameter gradients is evaluated. For any two loss terms $L_a$ and $L_b$, the gradient cosine similarity is defined as:

$$\cos\left(\nabla_\theta L_a, \nabla_\theta L_b\right) = \frac{\nabla_\theta L_a \cdot \nabla_\theta L_b}{\| \nabla_\theta L_a \|_2 \; \| \nabla_\theta L_b \|_2} \tag{42}$$

where $\theta$ denotes the trainable network parameters. A value close to 1 indicates strong alignment between the two gradients, a value close to -1 indicates direct opposition, and a value close to 0 indicates approximate orthogonality. In this study, the metric is used to examine the interaction among the PDE, boundary-condition, and initial-condition loss terms during training.

**(a)**

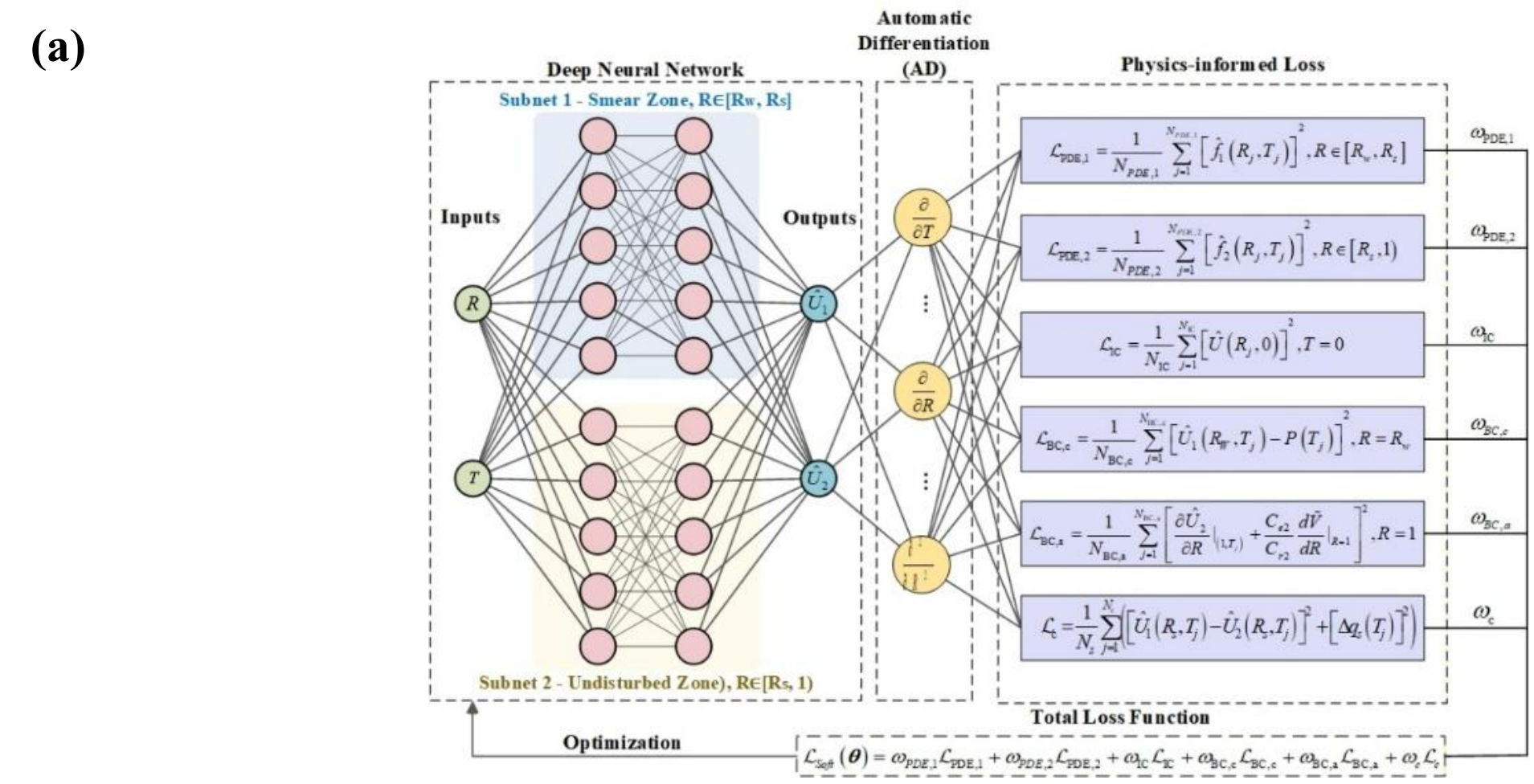

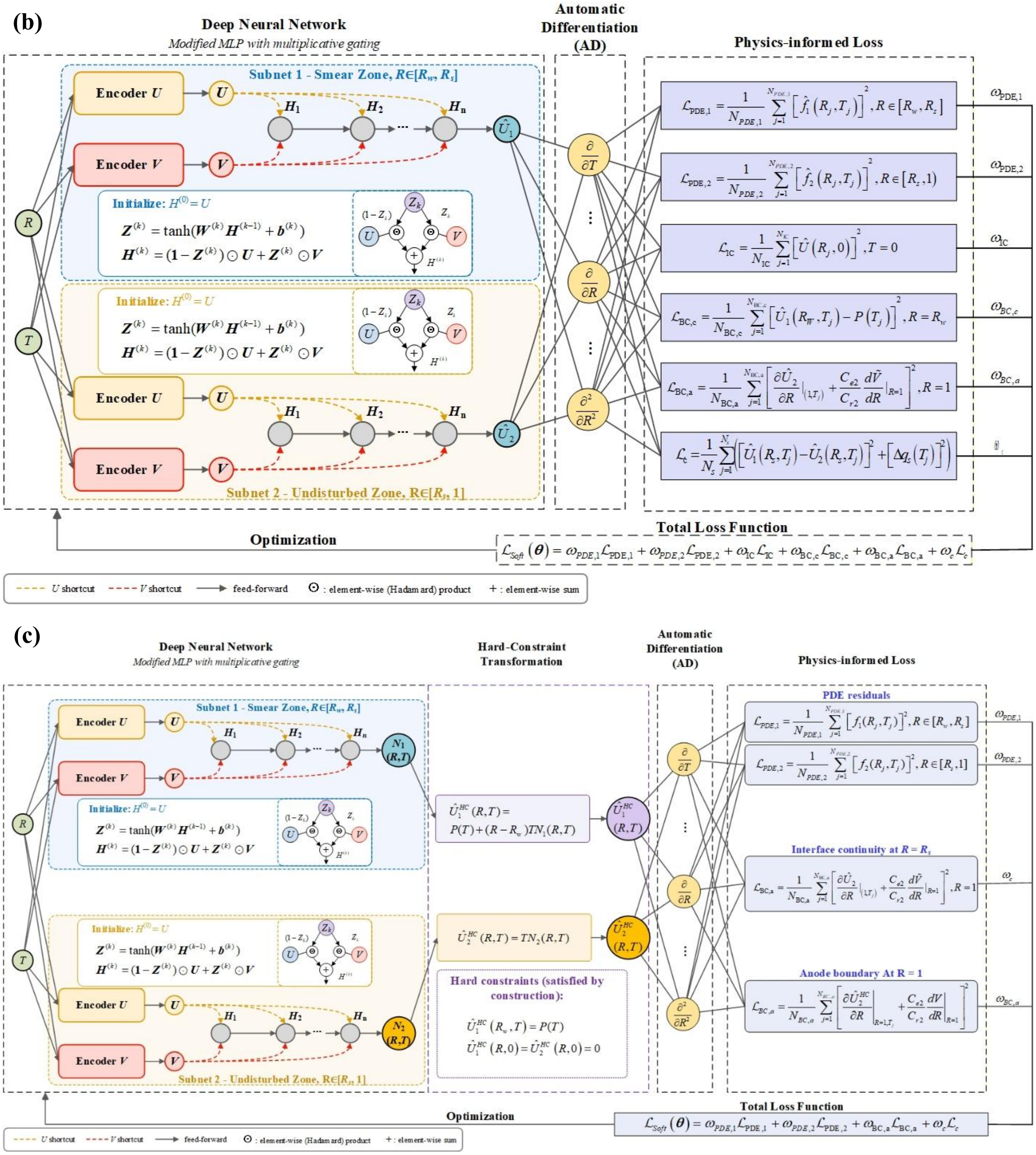

**Fig. 3.** Schematic diagrams of the three PINN frameworks: (a) standard PINN (Std-PINN), (b) modified MLP PINN (Mod-PINN), and (c) modified MLP PINN with hard constraints (Mod-HC-PINN).

## 3.6 Collocation and sampling strategy

PDE collocation points were sampled in the ($R$, $T$) domain using Latin Hypercube Sampling (LHS), which provides a stratified, space-filling distribution and more uniform coverage than purely random sampling (McKay et al., 1979). The fresh LHS points were regenerated each epoch to promote stochastic exploration, with $N_1 = 10{,}000$ interior points in the smear zone and $N_2 = 20{,}000$ in the undisturbed zone; the 2:1 allocation reflects the substantially larger radial span of Zone 2 ($1 - R_s = 0.667$) relative to Zone 1 ($R_s - R_w = 0.222$). Initial-condition enforcement used $N_{\mathrm{IC}} = 2{,}500$ uniformly spaced radial points along $T = 0$ over $R \in [R_w, 1]$. Boundary and interface constraints used 5,000 uniformly spaced time points over $T \in [0,1]$, evaluated at $R = R_w$ (cathode Dirichlet), $R = 1$ (anode Neumann), and $R = R_s$ (interface continuity).

# 4 Implementation

Three PINN models are evaluated in this study including Std-PINN, Mod-PINN, and Mod-HC-PINN. Each sub-network has seven hidden layers with 64 neurons per layer. The hyperbolic tangent (tanh) is chosen as the activation function for all hidden layers (Raissi et al., 2019). All network parameters are initialized using PyTorch's default Kaiming uniform scheme (He et al., 2015). For Cases C1-C3, the composite loss function is minimized using the Adam optimizer (Kingma and Ba, 2015) with an initial learning rate of $10^{-3}$ and a stepwise decay schedule (a factor of 0.5 every 5,000 epochs) over 25,000-30,000 epochs. For Case C4 (cyclic haversine loading), a two-stage strategy is adopted: 2,000 epochs of Adam warm-up followed by L-BFGS quasi-Newton optimization (history size 50, maximum 50,000 function evaluations, gradient tolerance $10^{-9}$). All neural network training was performed in Python 3.11.10 with PyTorch 2.x on an NVIDIA V100 (PCIe; 32 GB HBM2, 5,120 CUDA cores, 640 Tensor Cores). A finite element (FEM) model was implemented in COMSOL to generate reference solutions for the excess pore-water pressure under the governing equations, boundary, and initial conditions described in Section 2. Unless otherwise noted, all three PINN results are compared against these reference FEM solutions.

Two evaluation indices, namely the mean absolute error (MAE) and the mean relative error (MRE), are used to quantify the discrepancy between the PINN predictions and the reference FEM solutions as follows:

$$MAE = \frac{1}{N}\sum_{k=1}^{N}|u_i(R_k,T_k) - \tilde{u}_i(R_k,T_k)| \tag{43}$$

$$MRE = \frac{1/N\sum_{k=1}^{N}|u_i(R_k,T_k) - \tilde{u}_i(R_k,T_k)|}{1/N\sum_{k=1}^{N}|\tilde{u}_i(R_k,T_k)|} * 100\% \tag{44}$$

where $u_i$ is the simulated value predicted by PINNs; $\tilde{u}_i$ is the simulated value of the reference FEM solution; $N$ is the sample number.

# 5. Results and Discussion

The physical parameters adopted follow the axisymmetric electro-osmotic consolidation benchmarks of Liu et al. (2022), Zong et al. (2024) and Zhang et al. (2024), and are summarized in Table 1. The unit-cell geometry uses a drain radius $r_w$ = 0.2 m, smear-zone outer radius $r_s$ = 0.6 m, and influence radius $r_e$ = 1.8 m giving smear and drain-spacing ratios of $s = r_s/r_w$ = 3 and $n = r_e/r_w$ = 9 that are representative of field-installed EVDs. Within the smear zone, hydraulic permeability is reduced relative to the natural soil ($k_{r1}/k_{r2}$ = 0.5) while electro-osmotic and volume-compressibility properties are kept identical in both zones. The applied voltage at the anode is $V_{\max}$= 20 V, and the total simulation time is $T_{\text{final}}$ = 1000 d in all cases.

The loading conditions considered are summarized in Table 2. Case C1 applies a constant vacuum pressure $p(t)$ = -30kPa at the cathode with no surcharge, isolating the baseline electro-osmotic-vacuum response. Case C2 replaces the constant vacuum with a time-dependent exponential vacuum $p(t) = -p_v(1 - e^{-\alpha t})$, where $p_v$ = 30kPa and rate parameter $\alpha$ = 0.01/d, again with no surcharge. Case C3 retains the exponential vacuum and adds a ramp-and-hold surcharge of magnitude $q_u$ = 50kPa applied linearly over $t_c$ = 100d and held constant thereafter. Case C4 retains the same exponential vacuum but replaces the ramp surcharge with a cyclic haversine surcharge of amplitude $q_u$ = 50kPa and period $t_0$ = 73.96d, producing approximately 13-14 complete loading cycles over the 1,000-day simulation period.

**Table 1.** Physical parameters used in all four cases.

| Symbol | Description | Value | Unit |
|---|---|---|---|
| $r_w$ | Drain (cathode) radius | 0.2 | m |
| $r_s$ | Smear-zone outer radius | 0.6 | m |
| $r_e$ | Influence (anode) radius | 1.8 | m |
| $\gamma_w$ | Unit weight of water | 10 | $kN/m^3$ |
| $V_{max}$ | Anode voltage | 20 | V |
| $k_{r1}$ | Hydraulic permeability, smear zone | $3.46 \times 10^{-5}$ | m/d |
| $k_{r2}$ | Hydraulic permeability, undisturbed zone | $6.91 \times 10^{-5}$ | m/d |
| $k_{e1}$ | Electro-osmotic permeability, smear zone | $3.46 \times 10^{-5}$ | $m^2$ / V d |
| $k_{e2}$ | Electro-osmotic permeability, undisturbed zone | $3.46 \times 10^{-5}$ | $m^2$ / V d |
| $m_{v1}$, $m_{v2}$ | Volume compressibility | $1.0 \times 10^{-3}$ | /kPa |
| $p_v$ | Stabilized vacuum pressure (Cases C1-C4) | 30 | kPa |
| $\alpha$ | Vacuum rate parameter (Cases C2-C4) | 0.01 | /d |
| $q_u$ | Surcharge magnitude (Cases C3-C4) | 50 | kPa |
| $t_c$ | Ramp loading duration (Case C3) | 100 | d |
| $t_0$ | Haversine loading period (Case C4) | 73.96 | d |
| $T_{final}$ | Total simulation time | 1000 | d |

**Note:** Loading parameters are listed only for the cases in which they are used.

**Table 2.** Test matrix for the PINN simulations

| Case | Loading condition | Models evaluated |
|---|---|---|
| C1 | Constant vacuum | Std-PINN, Mod-PINN |
| C2 | Exponential vacuum | Std-PINN, Mod-PINN, Mod-HC-PINN |
| C3 | Exponential vacuum + ramp surcharge | Std-PINN, Mod-PINN, Mod-HC-PINN |
| C4 | Exponential vacuum + cyclic haversine surcharge | Std-PINN, Mod-PINN, Mod-HC-PINN |

**Note.** The Mod-HC-PINN is not applied to C1 because the constant vacuum boundary is incompatible with the zero initial condition at the cathode corner. The hard-constraint formulation is therefore reserved for C2-C4, where the exponential vacuum satisfies the initial condition at T = 0.

## 5.1. EO consolidation with constant vacuum

Fig. 4 compares the radial profiles of excess pore-water pressure $u(r)$ predicted by the Std-PINN and Mod-PINN against the reference FEM solution at selected times for Case C1. As shown in Fig. 4 (a), noticeable discrepancies between the Std-PINN and FEM profiles are observed, particularly at early to intermediate times. At t = 100 d, the Std-PINN predicted profile deviates from FEM reference over the radial domain. The gap narrows at later times but remains visible at t = 400 d. These discrepancies are consistent with the spectral bias of standard fully connected networks, where the sequential architecture preferentially learns the low-frequency (large-scale) features of the solution while under-resolving the sharper gradients near the cathode boundary and across the smear-zone transition. In contrast, the Mod-PINN results in Fig. 4(b) agree closely with the FEM reference at all six selected time instants, including the region near the cathode boundary where the steepest gradients occur. The gradient change at the smear-zone boundary ($r_s$ = 0.6 m), arising from the permeability contrast $k_{r2}/k_{r1} = 2$, is resolved smoothly by both sub-networks without visible discontinuity, confirming that the interface coupling via the continuity loss terms is effective.

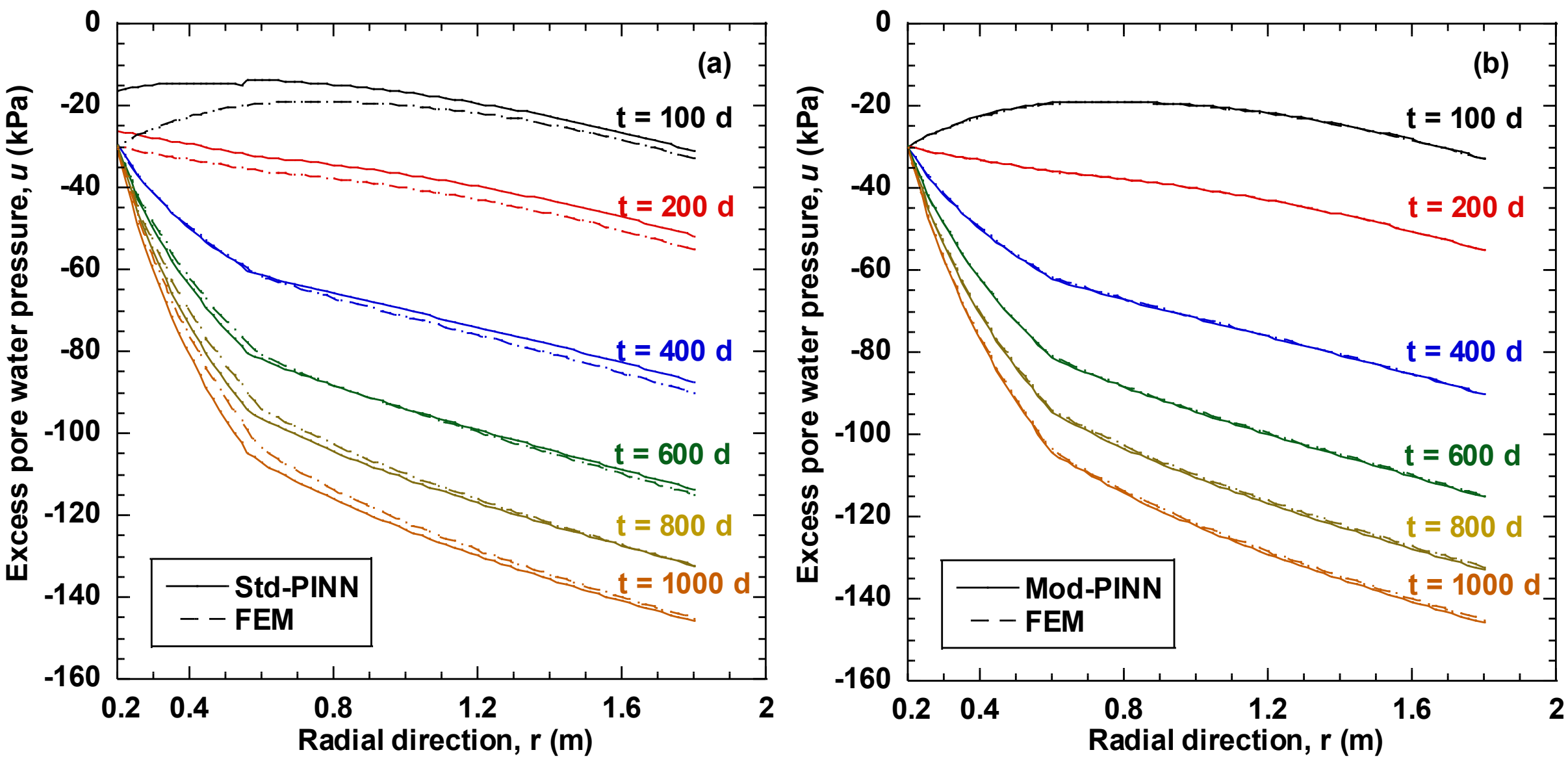

**Fig. 4.** Comparison of radial excess pore-water pressure profiles at selected times for Case C1: (a) Std-PINN versus FEM; (b) Mod-PINN versus FEM.

Fig. 5 presents the spatiotemporal contour maps of excess pore-water pressure for Case C1. The contour distributions indicate that both PINN models reproduce the overall consolidation behavior obtained from the FEM reference solution. The excess pore-water pressure becomes increasingly negative with radial distance from the cathode and with elapsed time, with the largest negative values developing near the outer boundary at late stages of consolidation. However, the corresponding absolute error maps [Fig. 5(d) and 5(e)] reveal difference in predictive performance between the two architectures. In the Std-PINN error map [Fig. 5(d)], the largest errors occur primarily in the smear zone and near the cathode at early times, where the radial pore-water pressure gradients are strongest. A second persistent error concentration develops around the smear-zone interface ($r \approx 0.6$ m), suggesting that the standard architecture is less effective in resolving the gradient transition associated with the permeability contrast between the smear zone and the surrounding undisturbed soil. In contrast, the Mod-PINN error field [Fig. 5(e)] is substantially reduced throughout the entire space-time domain, with absolute errors generally remaining below approximately 1 kPa. The remaining discrepancies are localized mainly near the cathode during the initial stage and within a narrow region adjacent to the interface at intermediate times, and in the outer undisturbed zone at late times. These results indicate that the modified gated architecture significantly improves both the local resolution of steep pore-pressure gradients and the overall prediction accuracy of the electro-osmotic consolidation response under constant vacuum loading. This interpretation is consistent with the quantitative metrics in Table 3, where the MAE decreases from 2.01 kPa to 0.32 kPa and the MRE decreases from 9.78% to 5.21%.

**Table 3.** Quantitative error metrics of PINN-predicted excess pore-water pressure relative to the FEM reference solution for all cases.

| Case | Architecture | MAE (kPa) | MRE (%) | $R^2$ |
|---|---|---|---|---|
| C1 | Std-PINN | 2.01 | 9.78 | 0.99 |
| | Mod-PINN | 0.32 | 5.21 | 1.00 |
| | Mod-HC-PINN | N/A | N/A | N/A |
| C2 | Std-PINN | 14.32 | 27.38 | 0.83 |
| | Mod-PINN | 0.81 | 6.03 | 1.00 |
| | Mod-HC-PINN | 0.43 | 5.79 | 1.00 |
| C3 | Std-PINN | 8.99 | 37.44 | 0.92 |
| | Mod-PINN | 2.15 | 9.77 | 1.00 |
| | Mod-HC-PINN | 0.41 | 5.89 | 1.00 |
| C4 | Std-PINN | 21.48 | 45.12 | 0.70 |
| | Mod-PINN | 4.68 | 11.95 | 0.98 |
| | Mod-HC-PINN | 0.27 | 5.49 | 1.00 |

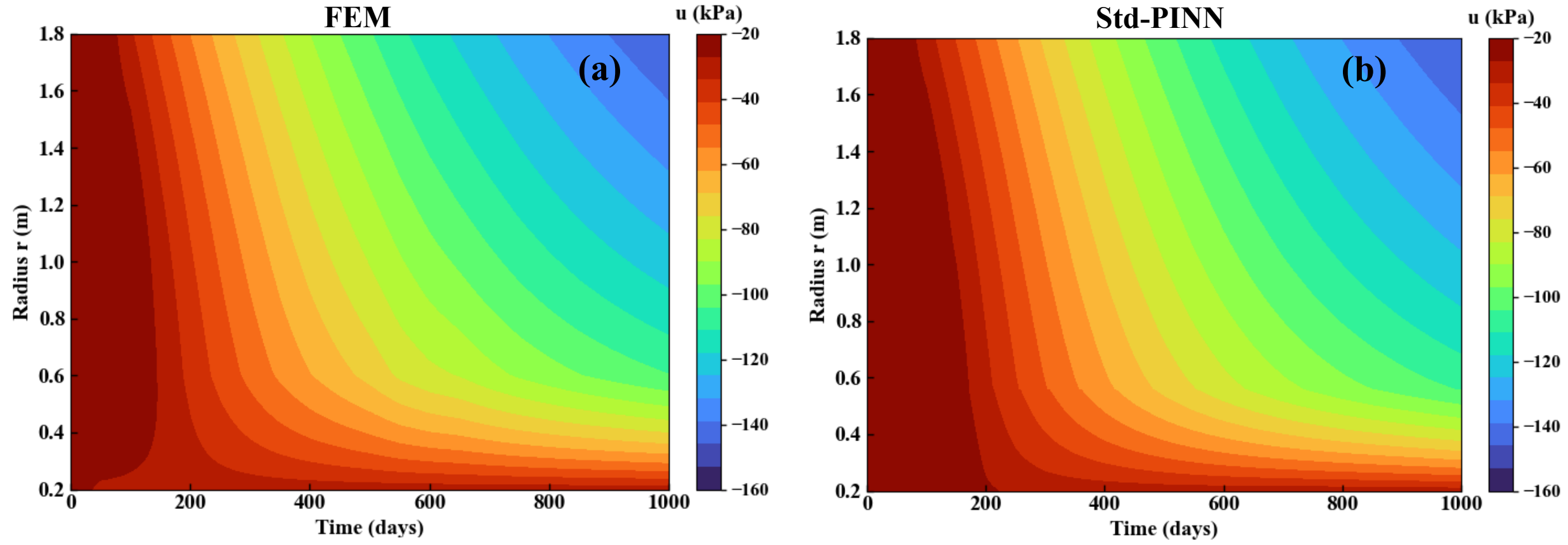

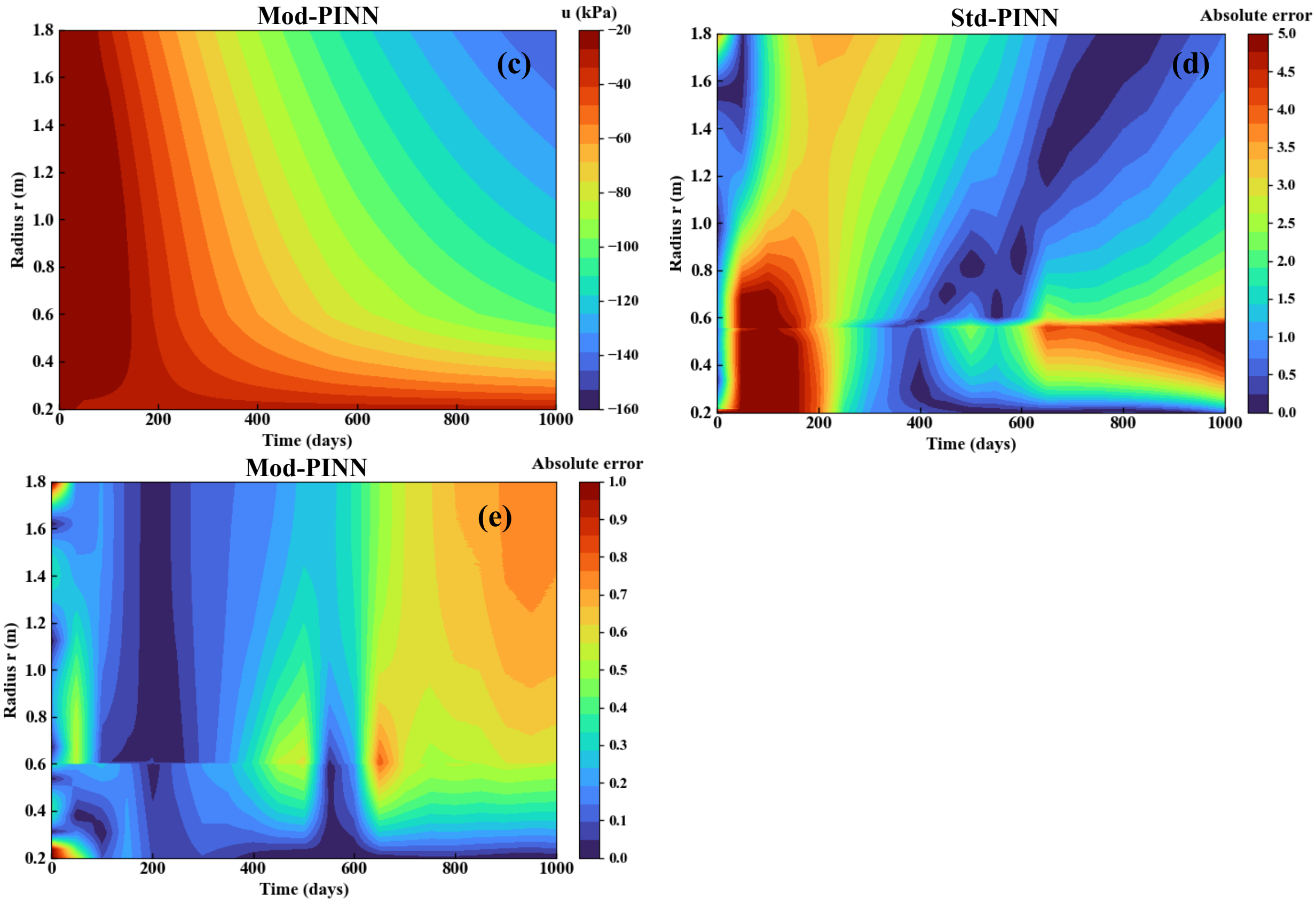

**Fig. 5.** Spatiotemporal distributions of excess pore-water pressure and corresponding absolute error fields for Case C1: (a) FEM reference solution; (b) Std-PINN prediction; (c) Mod-PINN prediction; (d) absolute error of the Std- PINN relative to the FEM solution. (e) absolute error of the Mod-PINN relative to the FEM solution.

Fig. 6 compares the training-loss histories of the Std-PINN and Mod-PINN models for Case C1. The Std-PINN total loss converges to approximately $10^{-3}$ after 15,000 epochs but continues to exhibit pronounced oscillations, particularly in the PDE-1 and PDE-2 components. These oscillations indicate that the standard architecture, lacking the multiplicative gating that improves NTK eigenvalue conditioning (Wang et al., 2022), cannot drive the PDE residuals to a stable floor within the training budget. In contrast, the Mod-PINN [Fig. 6(b)] shows smoother and more stable convergence throughout training, and the final total loss further decreases to the order of $10^{-4} \sim 10^{-3}$ by the end of training. The PDE-2 and CC (interface continuity) losses are reduced to the order of $10^{-5}$ to $10^{-6}$, while the IC and BC losses also converge more smoothly than in the Std-PINN. These results indicate that the modified gated architecture improves optimization stability and enables tighter overall enforcement of the governing equation, initial condition, boundary condition, and interface continuity constraint. This observation is consistent with the improved neural tangent kernel (NTK) eigenvalue conditioning reported by Wang et al. (2021).

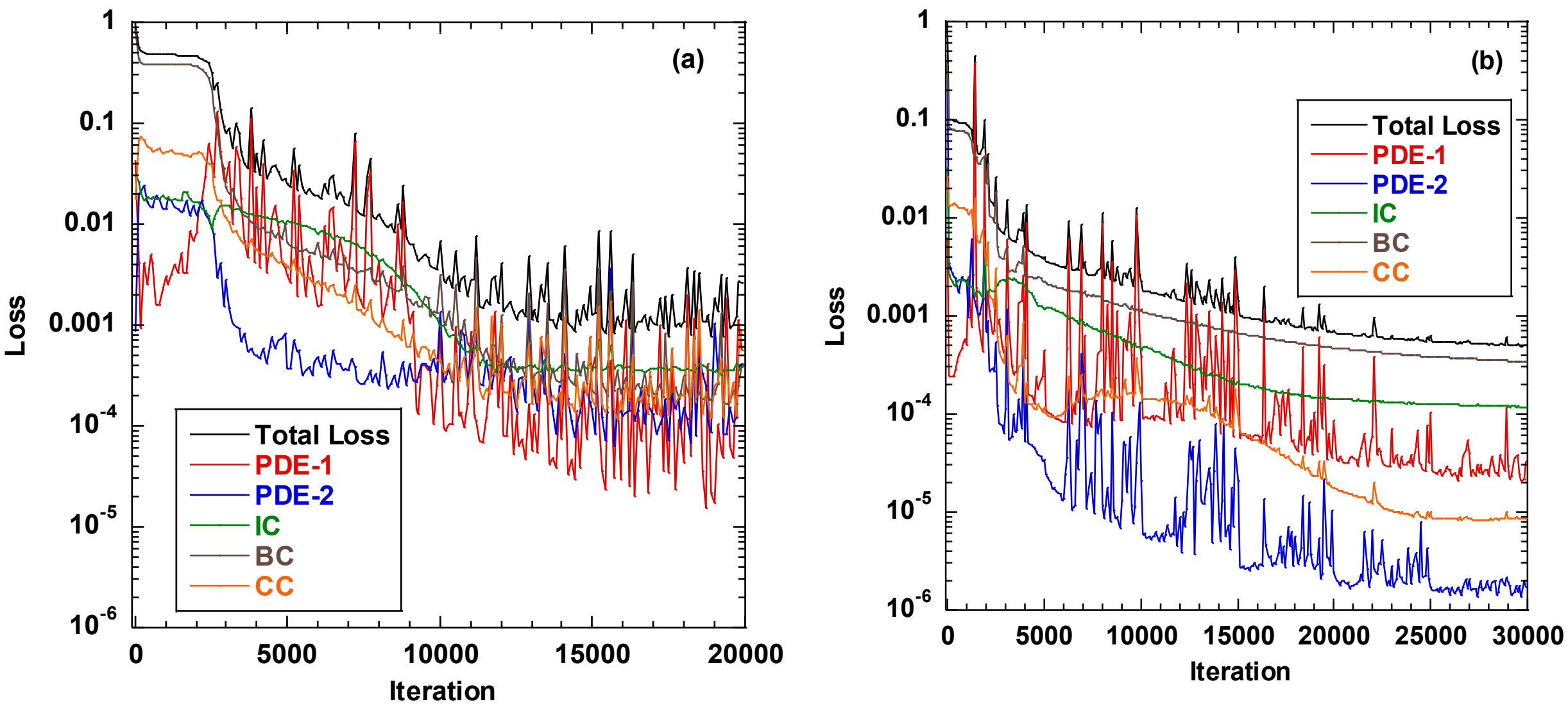

**Fig. 6.** Convergence histories of the total loss and individual loss components for the soft-constrained PINN models in Case C1: (a) Std-PINN; (b) Mod-PINN.

For the standard sequential MLP architecture, each hidden layer depends solely on the output of the previous layer. This sequential structure is more susceptible to spectral bias and therefore tends to learn the smooth, low-frequency component of the solution more readily than localized high-gradient features (Rahaman et al., 2019; Wang et al., 2022). In Case C1, this limitation is expressed mainly near the cathode and around the smear-zone interface, where the excess pore-water pressure varies most rapidly in space. The Mod-PINN alleviates this limitation through its multiplicative gating mechanism (Wang et al., 2021). Its dual encoder branches ($U$ and $V$) generate two independent nonlinear representations of the input, while the element-wise gated blending at each hidden layer provides richer feature combinations and more effective gradient propagation through the network. As a result, the modified architecture is better in resolving the steep spatial variation of excess pore-water pressure to achieve smoother and more stable optimization during training.

## 5.2. EO consolidation with exponential vacuum

Case C2 replaces the constant vacuum loading in Case C1 with an exponential vacuum history, $p(t) = -p_v(1 - e^{-\alpha t})$, where $p_v$ = 30 kPa and α = 0.01 day⁻¹. Accordingly, the applied vacuum pressure increases gradually from zero, reaching approximately 63% of $p_v$ at t = 100 d and 95% at t = 300 d. All other soil parameters, boundary conditions, and training settings remain unchanged. Fig. 7 compares the radial profiles of excess pore-water pressure u(r) predicted by the Std-PINN, Mod-PINN, and Mod-HC-PINN models at six selected time instants, alongside the FEM reference solution. As shown in Fig. 7(a), the Std-PINN exhibits large and systematic discrepancies from the FEM solution. At $t = 100$d, the Std-PINN substantially underpredicts the magnitude of excess pore-water pressure across the radial domain, particularly near the cathode, indicating that the Std-PINN fails to capture the early-stage transient response induced by the time-dependent vacuum loading. The discrepancies are not confined to the near-cathode region but extend across the full radial span, confirming that boundary errors propagate into the interior through the PDE. In contrast, the Mod-PINN reproduces the overall profile shape much more accurately, and the agreement with the FEM reference is substantially improved at all six selected time instants as shown in Fig. 7(b). However, the Mod-PINN profiles t = 400 ~ 1000 d remain visibly offset from the FEM reference. It indicates that, although the multiplicative gating mechanism improves the resolution of steep spatial gradients under static loading, it remains insufficient on its own to capture the full time-dependent evolution of $P(T)$ imposed along the cathode boundary. The Mod-HC-PINN predicted profiles agree closely with the FEM reference at all six selected time instants, including the intermediate and late stages of consolidation (t = 400-1000 d) where both the Std-PINN and Mod-PINN show visible deviations from the FEM reference (as shown in Fig. 7(c)). By embedding the cathode boundary condition and initial condition directly into the output transform (Eqs. 30–31), the Mod-HC-PINN satisfies these constraints exactly by construction. This reduces the optimization burden on boundary-loss enforcement and allows training to focus on the PDE residual, anode Neumann condition, and interface continuity constraints.

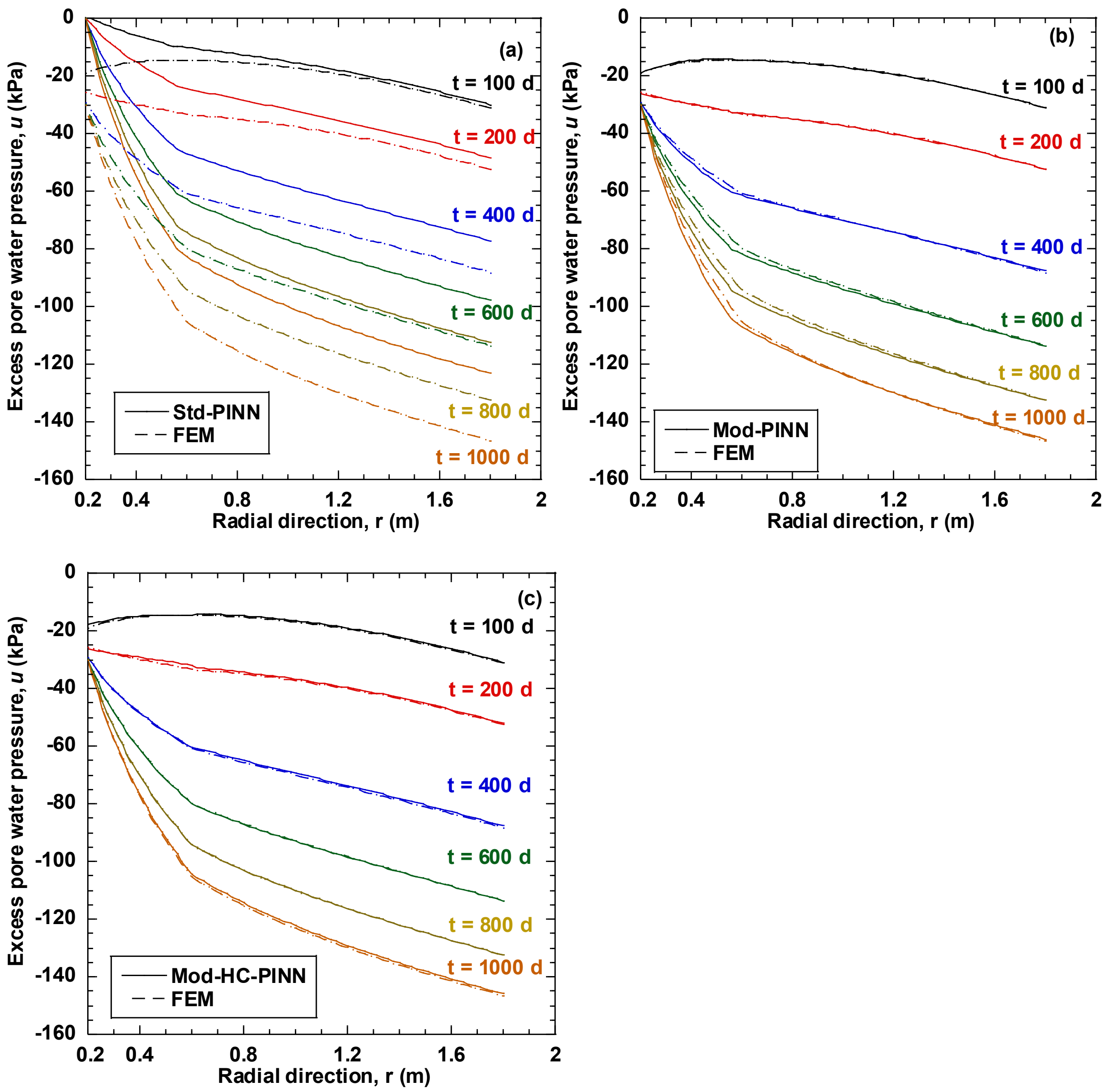

**Fig. 7.** Radial profiles of excess pore-water pressure at selected times for Case C2: (a) Std- PINN versus FEM; (b) Mod- PINN versus FEM; (c) Mod-HC-PINN versus FEM.

The spatiotemporal absolute error maps in Fig. 8 further clarify the performance difference among the three PINN models. The Std-PINN exhibits domain-wide errors that increase progressively with time, with peak values approaching 30 kPa (Fig. 8a). The Mod-PINN substantially reduces the peak error to approximately 4 kPa, but a pronounced error band remains concentrated near the smear-zone interface ($r \approx 0.6$ m) during intermediate to late times ($t \approx 400$–$1000$ d), together with smaller localized discrepancies near the cathode at early times (Fig. 8 (b)). These results indicate that, although the soft-constraint model captures the boundary and initial conditions more accurate than the Std-PINN, it still lacks sufficient capacity to simultaneously resolve the interior PDE residual and the interface flux continuity under time-dependent loading. The Mod-HC-PINN further reduces the error level, with peak errors limited to approximately 2 kPa, and confines the remaining discrepancies to localized regions near the smear-zone interface and during the early and late stages of consolidation (Fig. 8(c)). The broad high-error band observed in the Mod-PINN is therefore largely eliminated. This improvement confirms that embedding the cathode boundary condition and the initial condition directly into the network output reduces the multi-objective burden on the optimizer and allows more effective allocation of network capacity to the PDE residual, anode boundary condition, and interface continuity. The quantitative metrics further confirm the progressive improvement across the three PINN models in Case C2. The Std-PINN yields an MAE of 14.32 kPa, and an MRE of 27.38%, whereas the Mod-PINN reduces these values to 0.81 kPa, and 6.03%, respectively, and the Mod-HC-PINN further improves them to 0.43 kPa, and 5.79%.

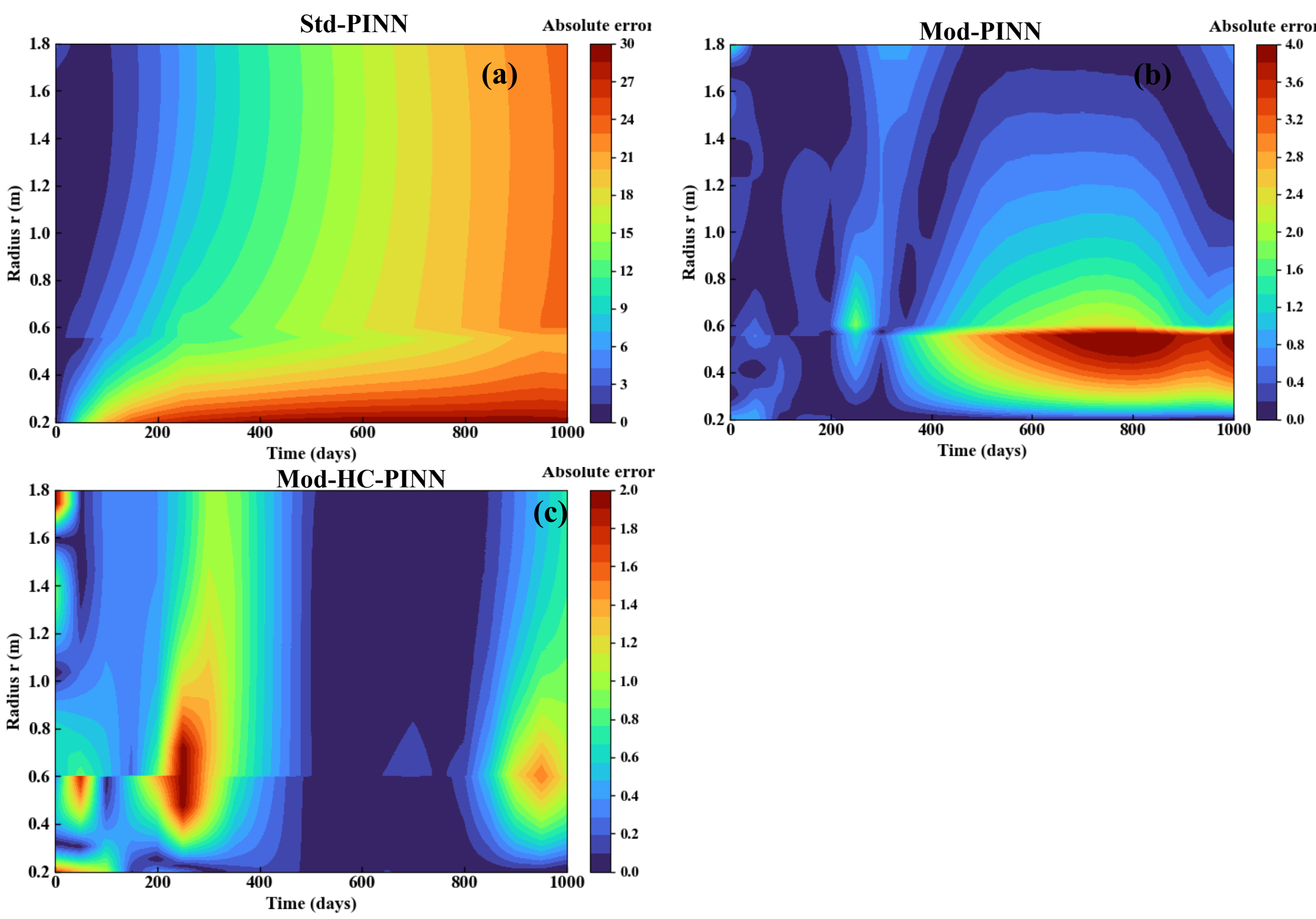

**Fig. 8.** Spatiotemporal absolute error distributions of (a) Std-PINN, (b) Mod-PINN, and (c) Mod-HC-PINN predictions for Case C2.

The training loss histories (Fig. 9) support the error-map observations. In the Mod-PINN [Fig. 9(a)], the PDE-1 and PDE-2 losses generally decrease with training, but the PDE-1 loss exhibits noticeable fluctuations and intermittent spikes. This suggests that the residual in the smear zone is more difficult to optimize, which is directly affected by the cathode boundary and interface conditions. Although the IC and BC losses are reduced to relatively low levels, the model must still balance the PDE residuals, initial condition, boundary conditions, and interface-continuity constraint within a single soft-constrained objective. This competition among loss components contributes to the localized errors observed near the cathode and the smear-zone interface. In the Mod-HC-PINN [Fig. 9(b)], because the initial condition and cathode Dirichlet boundary condition are imposed exactly by construction, their corresponding loss terms are eliminated from the optimization objective, leading to smoother convergence of the remaining PDE, anode BC, and interface continuity components. The total loss is also reduced to a lower level than that of the Mod-PINN. These results indicate that, once the initial condition and cathode boundary condition are imposed exactly by construction, the optimizer can focus more effectively on reducing the governing-equation residuals and the interface mismatch.

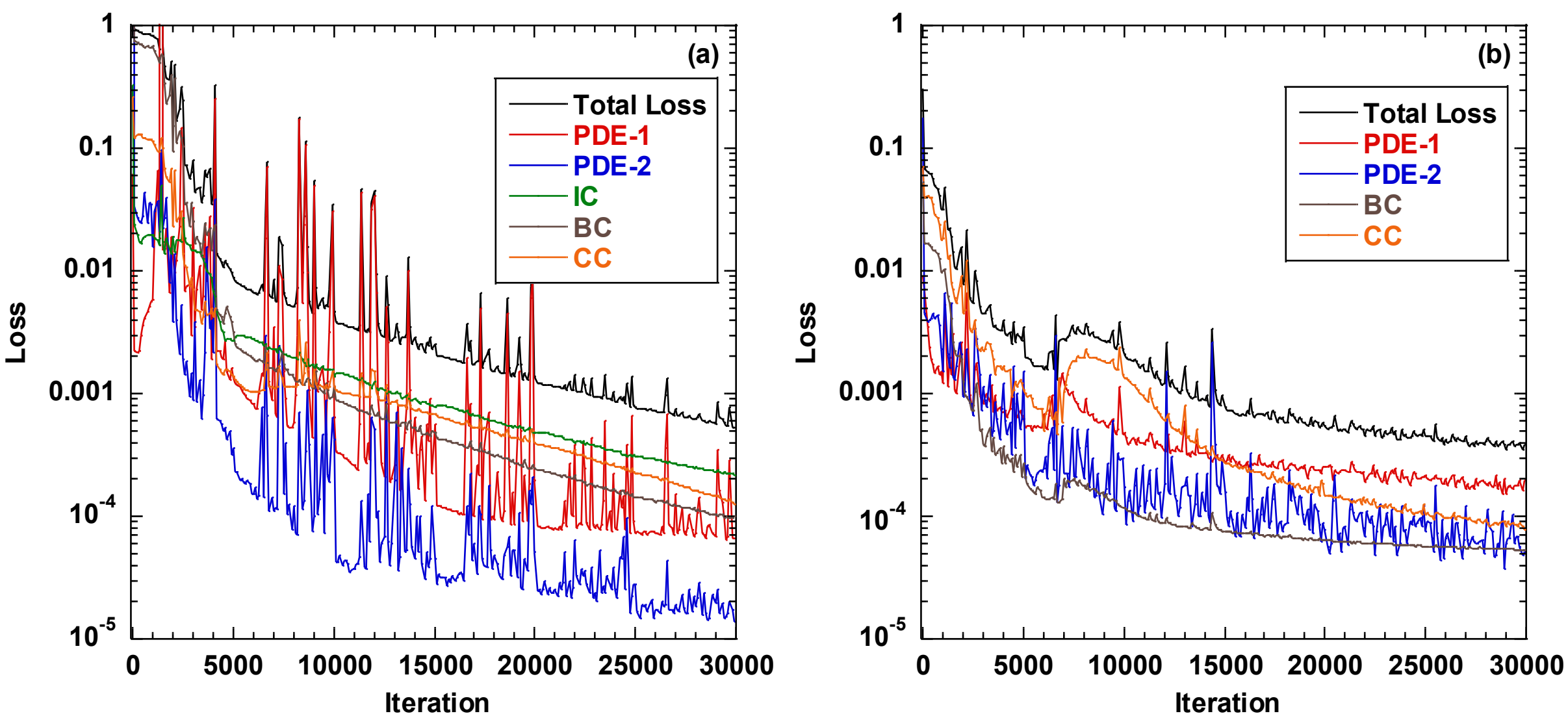

**Fig. 9.** Convergence histories of the total loss and individual loss components for the soft-constrained PINN models in Case C2: (a) Mod-PINN; (b) Mod-HC-PINN.

To examine the gradient dynamics associated with the different loss components, the cosine similarity between the PDE, boundary-condition (BC), and initial-condition (IC) loss terms gradients was computed every 500 iterations during Mod-PINN training for both Case C1 and Case C2 (Fig. 10). In Case C1 [Fig. 10(a)], $\cos(\nabla_\theta L_{BC}, \nabla_\theta L_{IC})$ progressively decreases and approaches -1.0 after approximately 17,000 iterations. It indicates that the BC and IC gradients become nearly antiparallel in late training stage, reflecting the incompatibility between the constant nonzero cathode boundary condition and the zero initial condition at the intersection point $(R_w, T = 0)$.

In Case C2 [Fig. 10(b)], $\cos(\nabla_\theta L_{BC}, \nabla_\theta L_{IC})$ does not decrease toward -1, but instead fluctuates around zero after the initial training stage. This occurs because the exponential vacuum satisfies P (0) = 0, making the cathode boundary condition compatible with the zero initial condition at the boundary–initial intersection point $(R_w, T = 0)$. In both cases, $\cos(\nabla_\theta L_{BC}, \nabla_\theta L_{PDE})$ and $\cos(\nabla_\theta L_{IC}, \nabla_\theta L_{PDE})$ remain close to zero in late training, indicating that the boundary-condition and initial-condition gradients are approximately orthogonal to the PDE-gradient direction.

It indicates that the reduced accuracy of the Mod-PINN in Case C2 is not primarily caused by direct gradient conflict between the boundary-condition and initial-condition losses. Instead, it reflects a multi-objective capacity limitation that the network must simultaneously learn the time-dependent cathode boundary function $P(T)$, propagate its effect into the interior domain, and minimize the PDE residual using the same finite set of trainable parameters. This interpretation is consistent with the loss histories in Fig. 9, where the boundary-condition and initial-condition losses in Case C2 continue to decrease during training rather than reaching the low floor values observed in Case C1. The hard-constraint transform introduced in Section 3.4 removes this capacity competition by embedding $P(T)$ directly into the output structure, allowing the trainable component of the network to focus on the PDE residual, anode boundary condition, and interface continuity constraints.

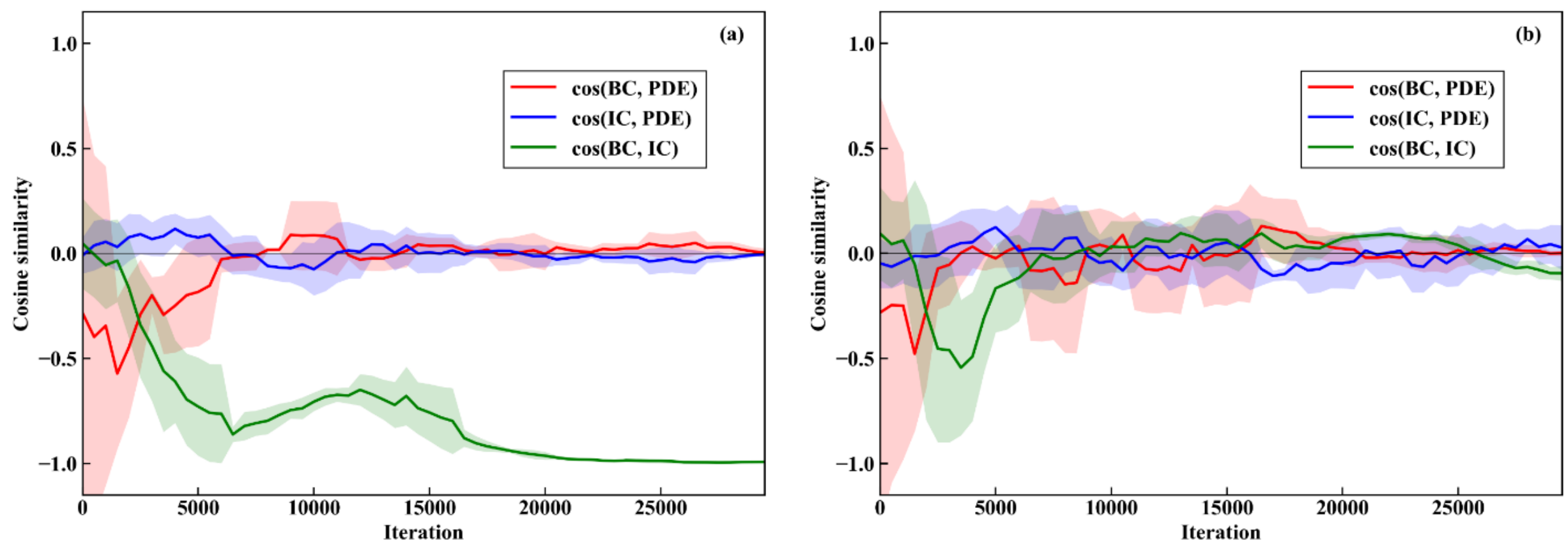

**Fig. 10.** Gradient cosine similarities between the PDE, BC, and IC loss terms during Mod-PINN training: (a) Case C1, constant vacuum; (b) Case C2, exponential vacuum.
Note: Lines show rolling mean; shaded bands show ±1 rolling standard deviation over a 2500-iteration window.

## 5.3 Combined electro-osmosis, vacuum, and ramp surcharge loading

Case C3 introduces a more complex consolidation response due to the combination of electro-osmosis, exponential vacuum loading, and ramp surcharge loading. As shown in Fig. 11(a), the Mod-PINN captures the general trend of the radial pore-water pressure distribution. The most noticeable discrepancies occur at later times, particularly from $t = 400$ to $1000\,\mathrm{d}$, when the excess pore-water pressure becomes fully negative. During this stage, the Mod-PINN generally predicts pressures that are less negative than the FEM reference, indicating that the long-term vacuum-dominated dissipation is still not captured with sufficient accuracy. In contrast, the Mod-HC-PINN [Fig. 11(b)] matches the FEM solution closely at all the six time instants, including the later consolidation stage.

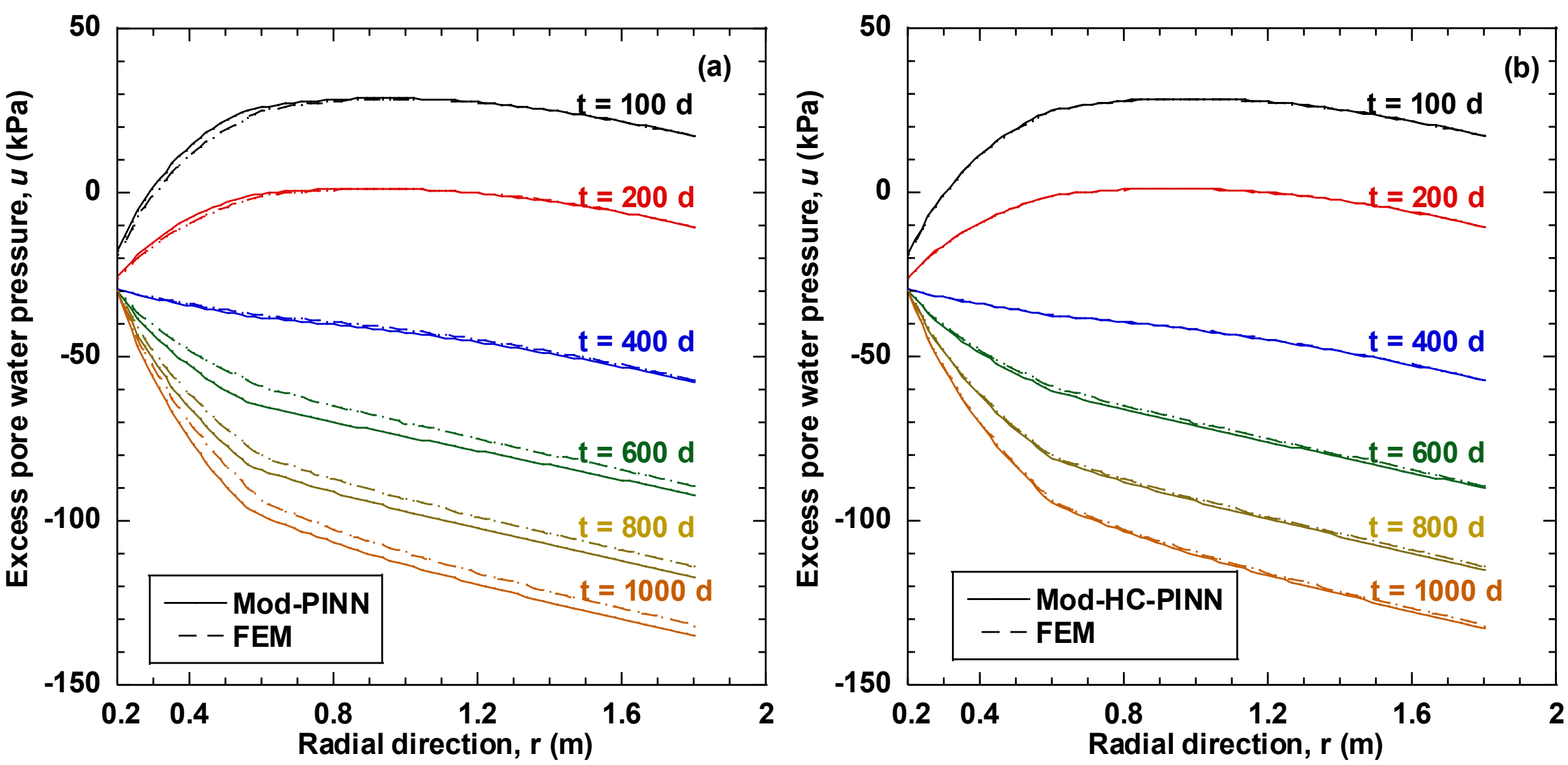

**Fig. 11.** Radial profiles of excess pore-water pressure at selected times for Case C3: (a) Mod- PINN versus FEM; (b) Mod-HC-PINN versus FEM.

The Mod-PINN [Fig. 12(a)] exhibits two main error features. The first is a localized early-time hotspot in the smear zone, associated with the surcharge-affected stage. The second is a strong intermediate-time hotspot centered near the smear-zone interface around $r \approx 0.6$ m and $t \approx 500$-$650$ d, with peak absolute errors of approximately 5 kPa. In addition, elevated errors persist along the smear-zone interface at later times, indicating that the soft-constraint model does not fully capture the subsequent vacuum-dominated dissipation response. The Mod-HC-PINN [Fig. 12(b)] substantially suppresses this pattern. The peak error is reduced to approximately 2 kPa and remains concentrated in a much smaller hotspot near the smear interface at intermediate times. The late-time error within the smear zone is also reduced. Thus, the hard-constraint formulation improves both the magnitude and the spatial confinement of the error under ramp surcharge loading. The quantitative metrics likewise show a clear improvement from Mod-PINN to Mod-HC-PINN in Case C3. The MAE decreases from 2.15 to 0.41 kPa, and the MRE from 9.77% to 5.89%.

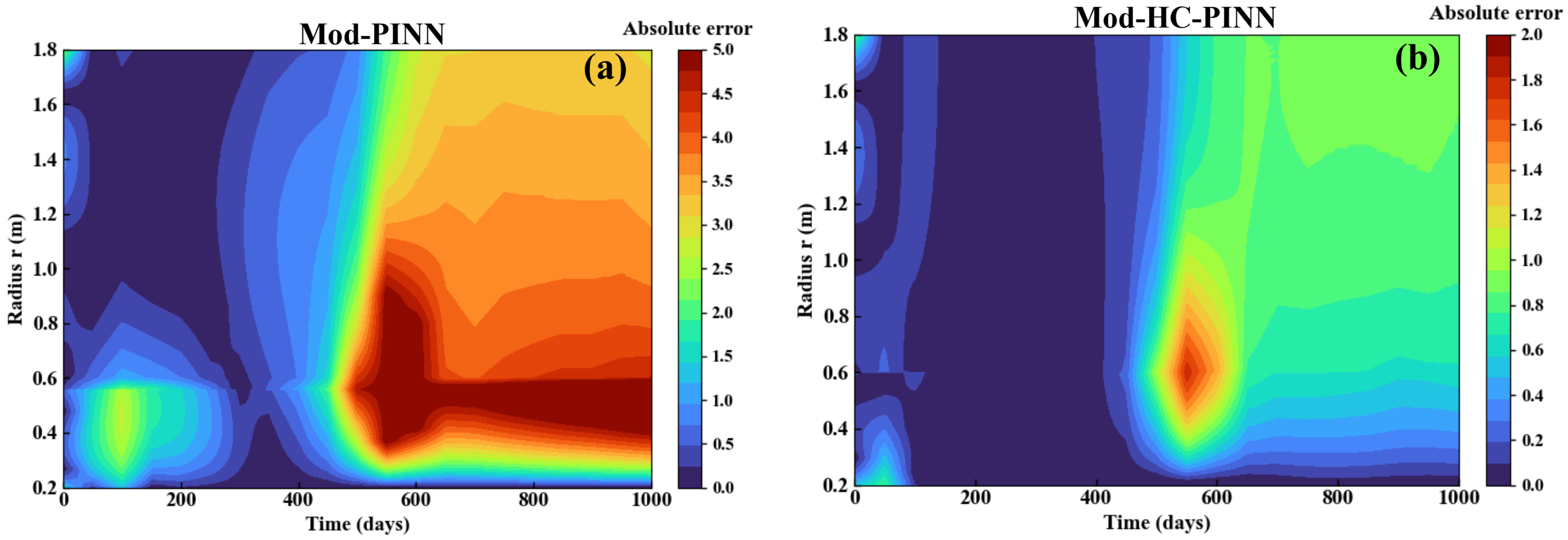

**Fig. 12.** Spatiotemporal absolute error distributions of (a) Mod-PINN, and (b) Mod-HC-PINN predictions for Case C3.

## 5.4 Combined electro-osmosis, vacuum, and cyclic haversine surcharge

Case C4 represents the most demanding configuration considered in this study. The surcharge follows a cyclic haversine function (as shown in Eq. (6)) superimposed on the exponential vacuum ramp, producing approximately 13-14 complete loading cycles over the 1,000-day simulation period. As shown in Fig. 13(a), the Mod-PINN model shows close agreement with the FEM reference during the early stages of loading (e.g., 100, 200 d) but exhibits systematic deviations from the FEM reference at the intermediate and late stages. At t = 600 d, a modest gap becomes visible for r > 1.0 m, with the Mod-PINN result profile lying slightly above the FEM reference. By t = 800 and 1,000 d, the discrepancy increases to approximately 10 kPa across the outer radial zone. This indicates that the soft-constraint model progressively underpredicts the cumulative vacuum-induced pore-pressure dissipation during repeated cyclic loading. The Mod-HC-PINN [Fig. 13(b)] matches the FEM solution closely at all the six selected time instants.

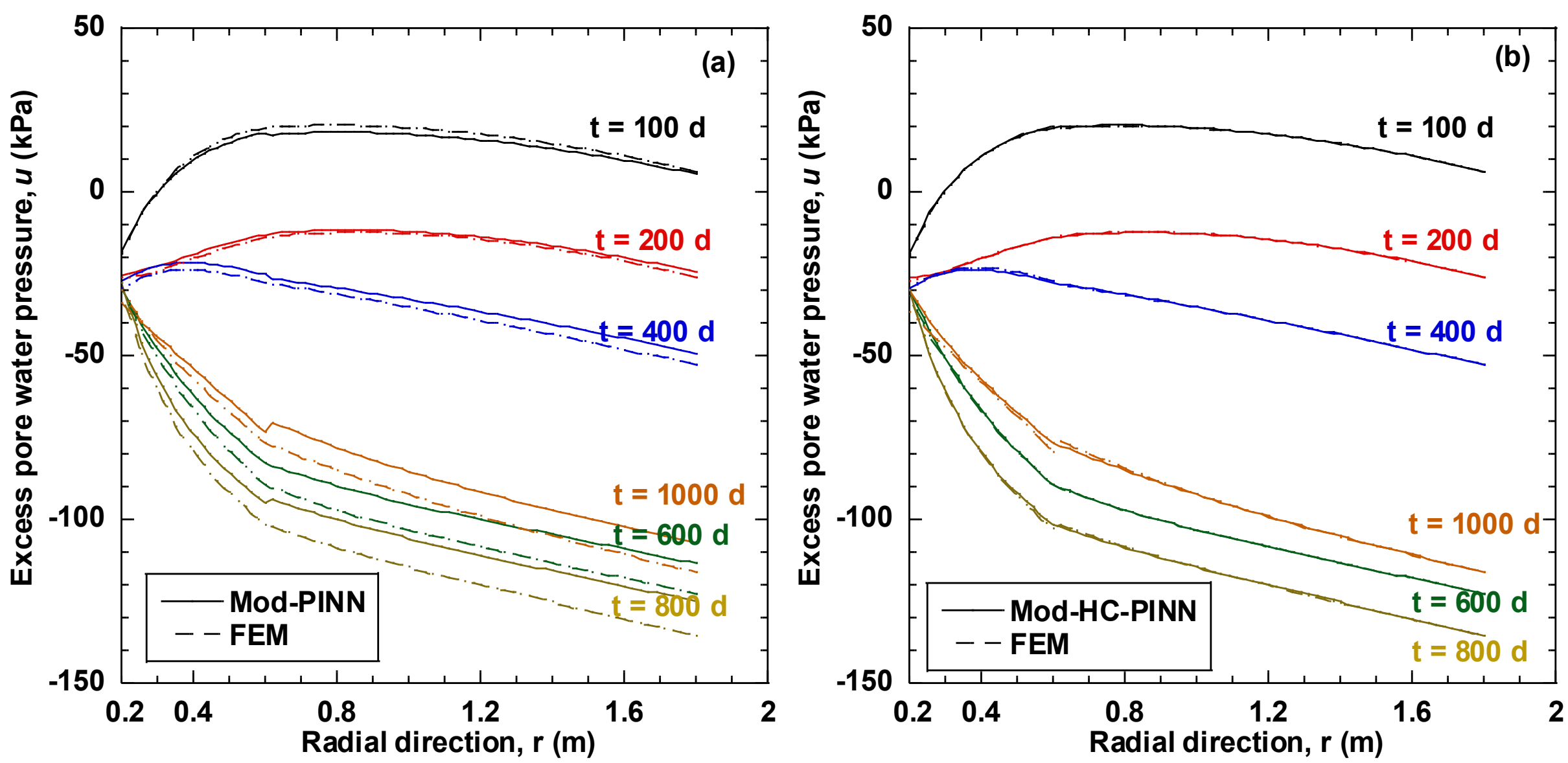

**Fig. 13.** Radial profiles of excess pore-water pressure at selected times for Case C4: (a) Mod- PINN versus FEM; (b) Mod-HC-PINN versus FEM.

Fig. 14 shows the time histories of excess pore-water pressure $u(t)$ at fixed radial locations r = 0.4, 0.6, 0.8, 1.2, and 1.6 m. For Mod-PINN [Fig. 14(a)], the Mod-PINN and FEM curves remain in reasonable agreement at the innermost location (r = 0.4 m) but diverge progressively outward. At r = 1.2 and 1.6 m, the cycle amplitudes are under-predicted, and a phase drift becomes apparent after approximately 600 d, indicating that the soft-constraint model does not adequately reproduce the cumulative cyclic pore-pressure response in the outer radial zone. For Mod-HC-PINN [Fig. 14(b)], the Mod-HC-PINN and FEM curves are nearly coincident at all radii and throughout the full 1,000-day simulation period.

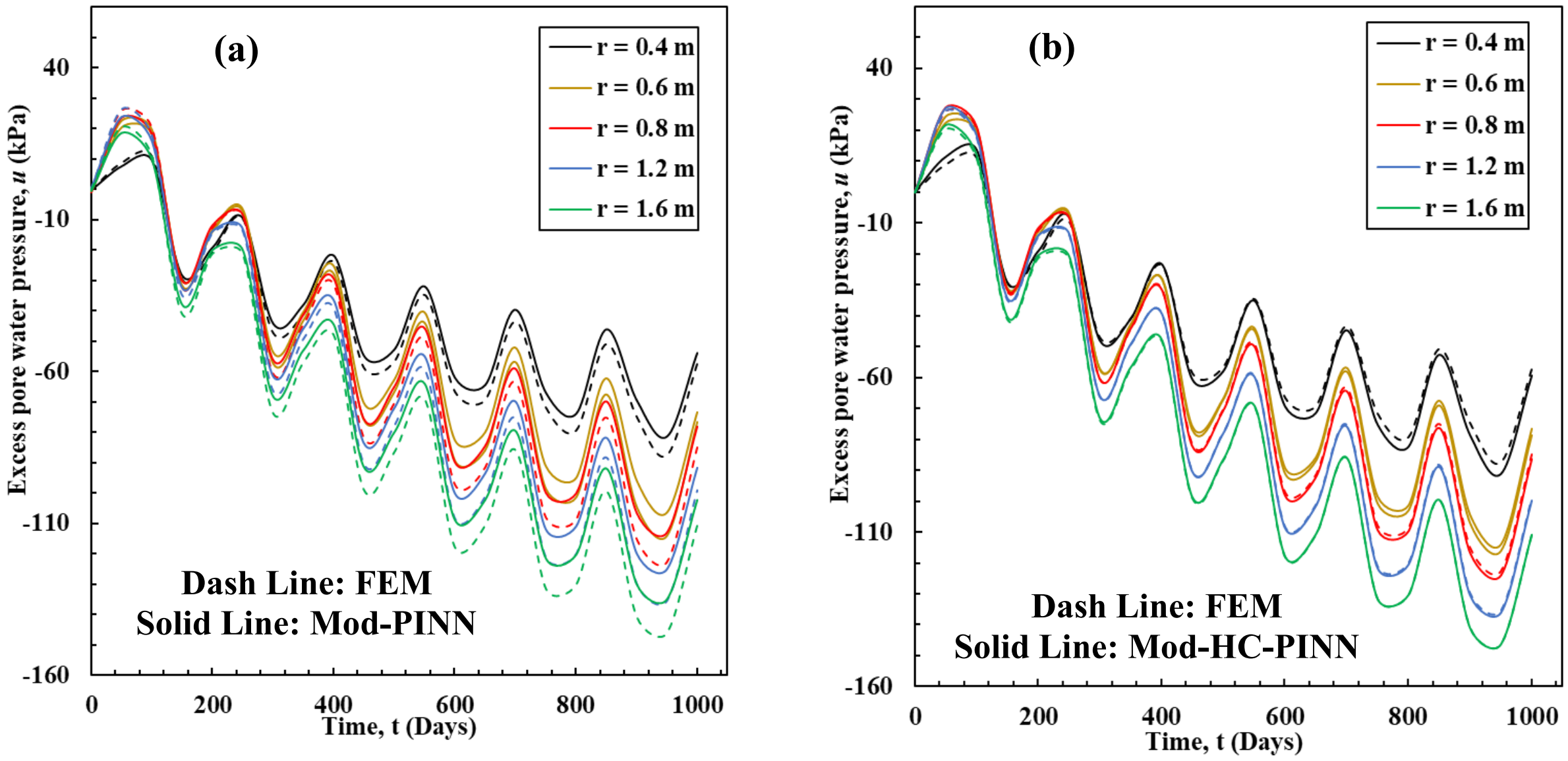

**Fig. 14.** Time histories of excess pore-water pressure at selected radial locations for Case C4: (a) Mod-PINN versus FEM; (b) Mod-HC-PINN versus FEM.

Fig. 15 provides spatiotemporal absolute error distributions over the (r, t) domain. The Mod-PINN error map [Fig. 15(a)] shows broad error accumulation in the undisturbed zone ($r > 0.6$ m), with peak absolute errors approaching 10 kPa. The errors form distinct periodic bands that follow the haversine surcharge cycles, indicating that the soft-constraint model does not fully resolve the repeated loading–unloading response. The error magnitude also increases with time, particularly in the outer radial region, indicating progressive error accumulation in resolving the coupled cyclic surcharge response and time-dependent vacuum-induced pore-pressure dissipation. For the Mod-HC-PINN map [Fig. 15(b)], the maximum error is reduced to below 5 kPa, and the elevated-error region is largely confined to localized areas at small radii and near the smear-zone interface. The undisturbed zone remains at comparatively low error levels, generally on the order of 1-2 kPa or less. The hard-constraint formulation therefore reduces the frequency-related error accumulation that characterizes the soft-constraint response. The quantitative metrics likewise show a clear improvement from Mod-PINN to Mod-HC-PINN in Case C4. The MAE decreases from 4.68 to 0.27 kPa, and the MRE from 11.95% to 5.49%.

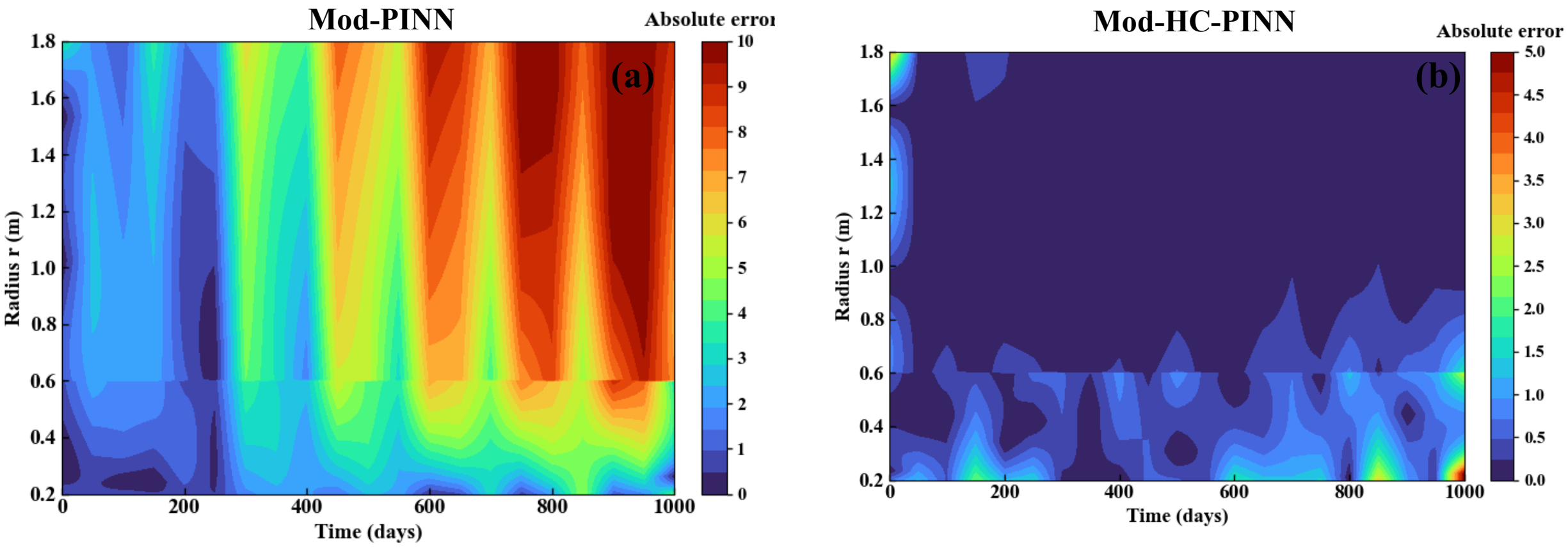

**Fig. 15.** Spatiotemporal absolute error distributions of (a) Mod-PINN, and (b) Mod-HC-PINN predictions for Case C4.

Case C4 demonstrates that cyclic surcharge loading intensifies the difficulty of the time-dependent boundary-value problem. In the soft-constraint Mod-PINN, the repeated oscillatory loading leads to phase drift and amplitude mismatch in the outer radial zone, resulting in periodic error banding in the spatiotemporal error map. By embedding $P(T)$ directly into the output transformation, the Mod-HC-PINN removes the need to learn the cathode boundary history from penalty terms and allows the trainable network component to focus on the PDE residual, anode boundary condition, and interface continuity. This substantially reduces the cumulative cyclic error and improves the agreement with the FEM reference solution.

## 5.5 Sensitivity analysis

### 5.5.1 Effect of structure of neural network

To evaluate the robustness of the Mod-HC-PINN framework, a systematic sensitivity analysis was performed by varying the network depth (1, 3, 5, 7, and 10 hidden layers) and width (8, 16, 32, 64, and 128 neurons per layer) for Case C3. Fig. 16 presents the mean absolute error (MAE) across all 25 configurations. The results reveal that both depth and width influence prediction accuracy, although network width exerts a stronger influence on accuracy than depth. Configurations with only 8 neurons per layer consistently yielded MAE values between 2.4 kPa and 10.4 kPa regardless of depth. Similarly, a single hidden layer failed to achieve acceptable accuracy (MAE = 2.36 kPa) even with 128 neurons, demonstrating that shallow architectures cannot adequately capture the spatial complexity of the proposed consolidation problem. The optimal performance region lies within 5-10 hidden layers and 32-128 neurons per layer, where MAE values remain below 0.6 kPa. The minimum MAE of 0.414 kPa was obtained for the 7-layer, 64-neuron configuration, though comparable accuracy was obtained across a broad range of configurations within the optimal region. In particular, similarly low errors are also achieved for nearby architectures, including 7 layers × 32 neurons (0.432 kPa), 10 layers × 32 neurons (0.415 kPa), and 7 layers × 128 neurons (0.418 kPa). Notably, increasing depth beyond 7 layers led to marginal degradation in accuracy, likely due to optimization difficulties in deeper networks. These findings confirm that the Mod-HC-PINN framework is robust to hyperparameter selection within a reasonable range and does not require extensive architecture tuning to achieve accurate predictions.

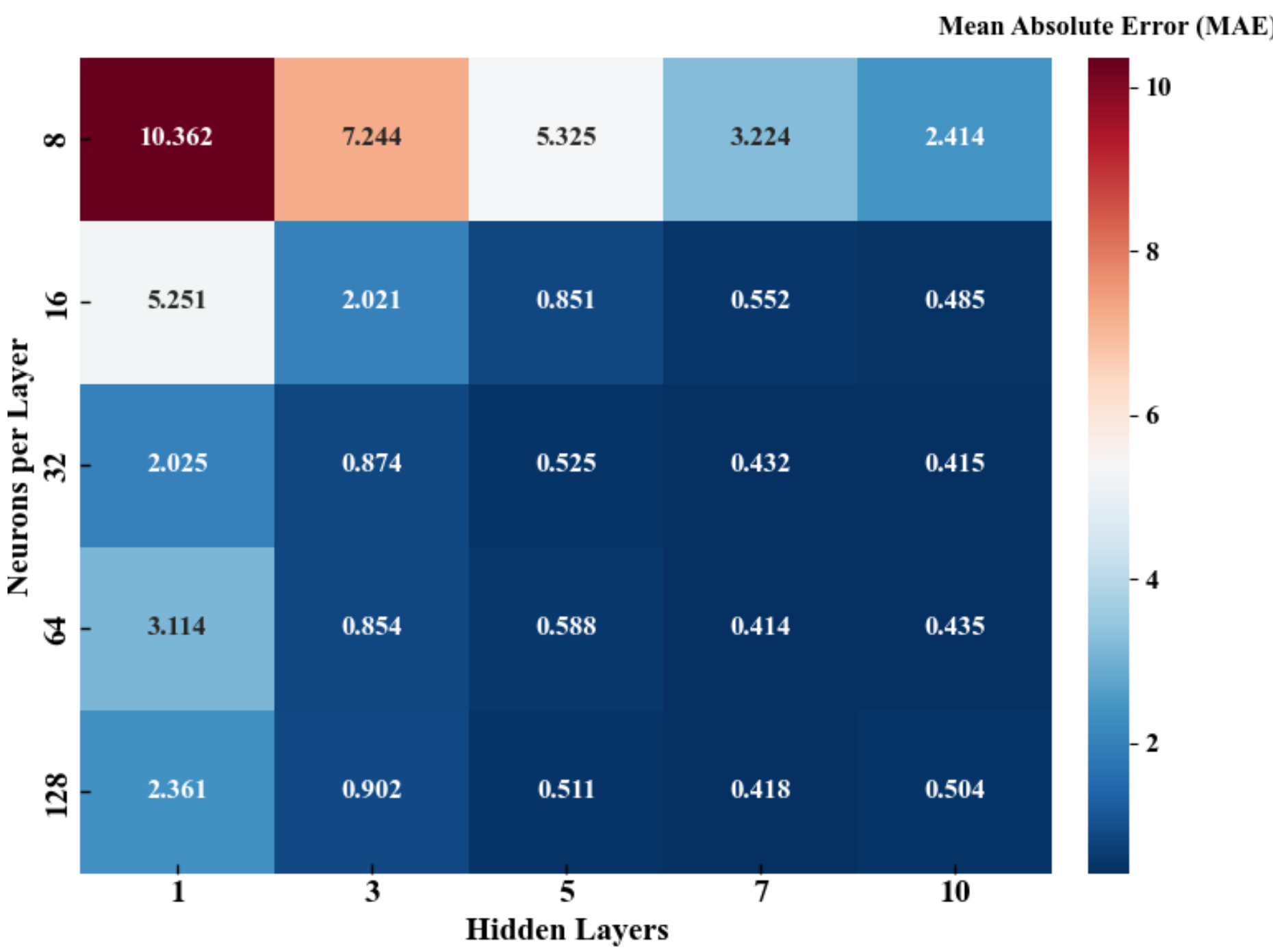

**Fig. 16.** Heatmap of MAE for different network architectures with varying numbers of hidden layers and neurons per layer.

### 5.5.2 Effect of training data

The influence of collocation point density on prediction accuracy was investigated by varying the total number of sampling points from 5,000 to 60,000 while maintaining the 1:2 ratio between smear zone and undisturbed zone. As shown in Fig. 17, the MAE decreases monotonically from 1.525 kPa at 5,000 points to 0.414 kPa at 30,000 points. Beyond this threshold, increasing the collocation points to 60,000 provides no further reduction in MAE, with the same value of 0.414 kPa obtained at both 30,000 and 60,000 points, indicating that the solution has converged with respect to spatial sampling. Configurations with fewer than 15,000 points show noticeably larger errors, indicating inadequate sampling density for accurate resolution of the governing residuals. These findings demonstrate that the Mod-HC-PINN framework achieves optimal accuracy at moderate computational cost without requiring excessive collocation point densities.

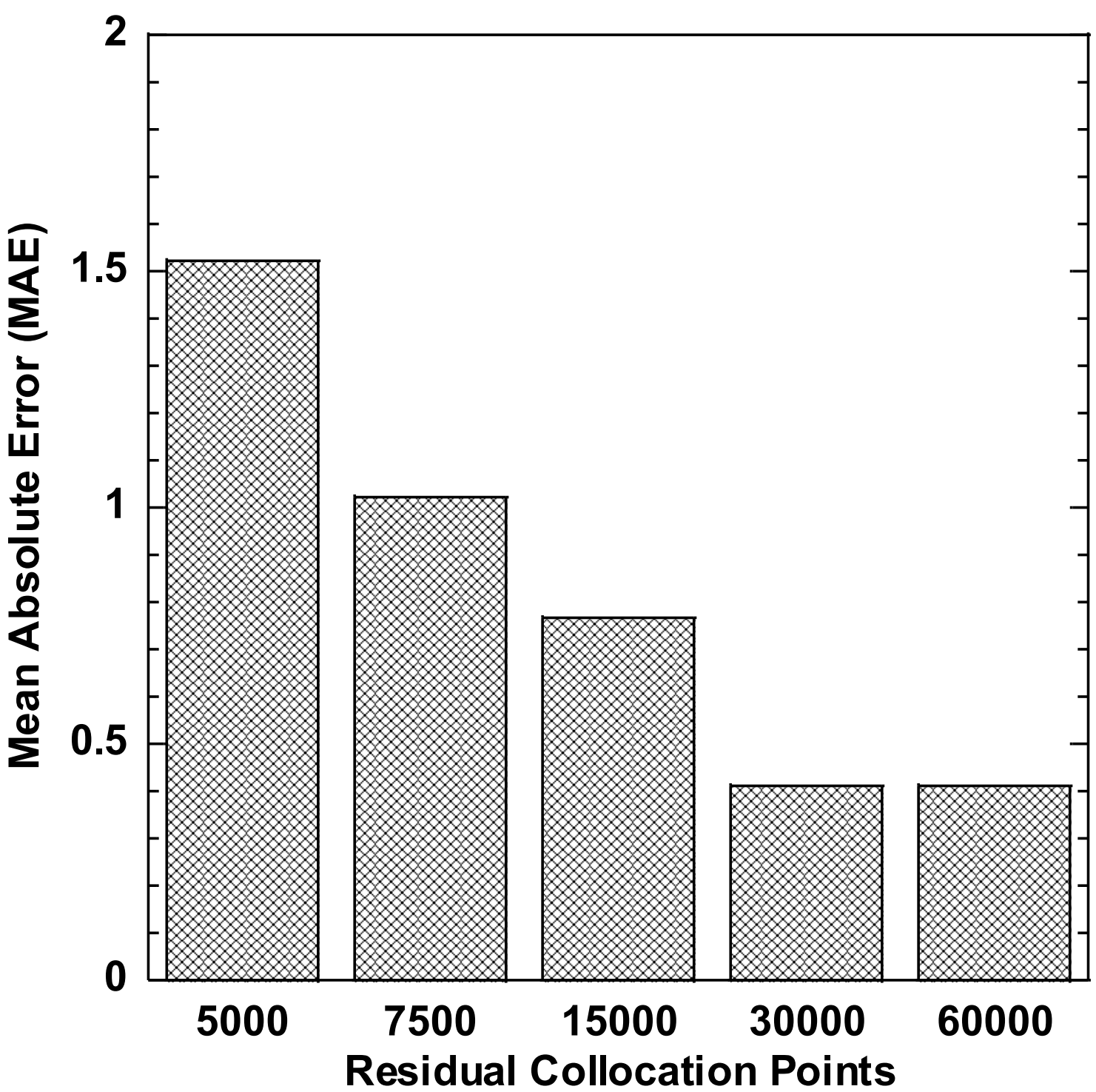

**Fig. 17.** Variation of mean absolute error (MAE) with the number of residual collocation points**.**

### 5.5.3 Effect of hydraulic permeability coefficients ratio on consolidation behavior

The effect of hydraulic permeability coefficients ratio on consolidation behavior was evaluated by varying the hydraulic permeability ratio $k_{r1}/k_{r2}$ from 0.1 to 1.0, corresponding to permeability reductions of 10× to no reduction, respectively. As shown in Fig. 18(a), the MAE increases from 0.247 kPa for homogeneous soil (ratio = 1.0) to 0.625 kPa at $k_{r1}/k_{r2} = 0.1$. This trend reflects the increasing difficulty of capturing the sharp pore-water pressure gradient at the smear zone interface as the permeability contrast intensifies. Nevertheless, the Mod-HC-PINN maintains MAE values below approximately 0.65 kPa even under the most challenging conditions, demonstrating its capability to handle realistic smear zone scenarios encountered in field applications.

### 5.5.4 Effect of electro-osmotic permeability coefficients ratio on consolidation behavior

The effect of electro-osmotic permeability ratio between the smear zone and undisturbed zone on consolidation behavior was assessed by varying $k_{e1}/k_{e2}$ from 0.1 to 1.0. The results show a similar trend to the hydraulic permeability analysis, with MAE decreasing from 0.524 kPa at ratio 0.1 to 0.414 kPa at ratio 1.0. The reduced electro-osmotic permeability in the smear zone creates additional complexity in the flux balance at the interface, which moderately increases prediction error. However, all configurations achieve MAE values below 0.55 kPa, confirming that the framework remains accurate across a wide range of electro-osmotic conditions.

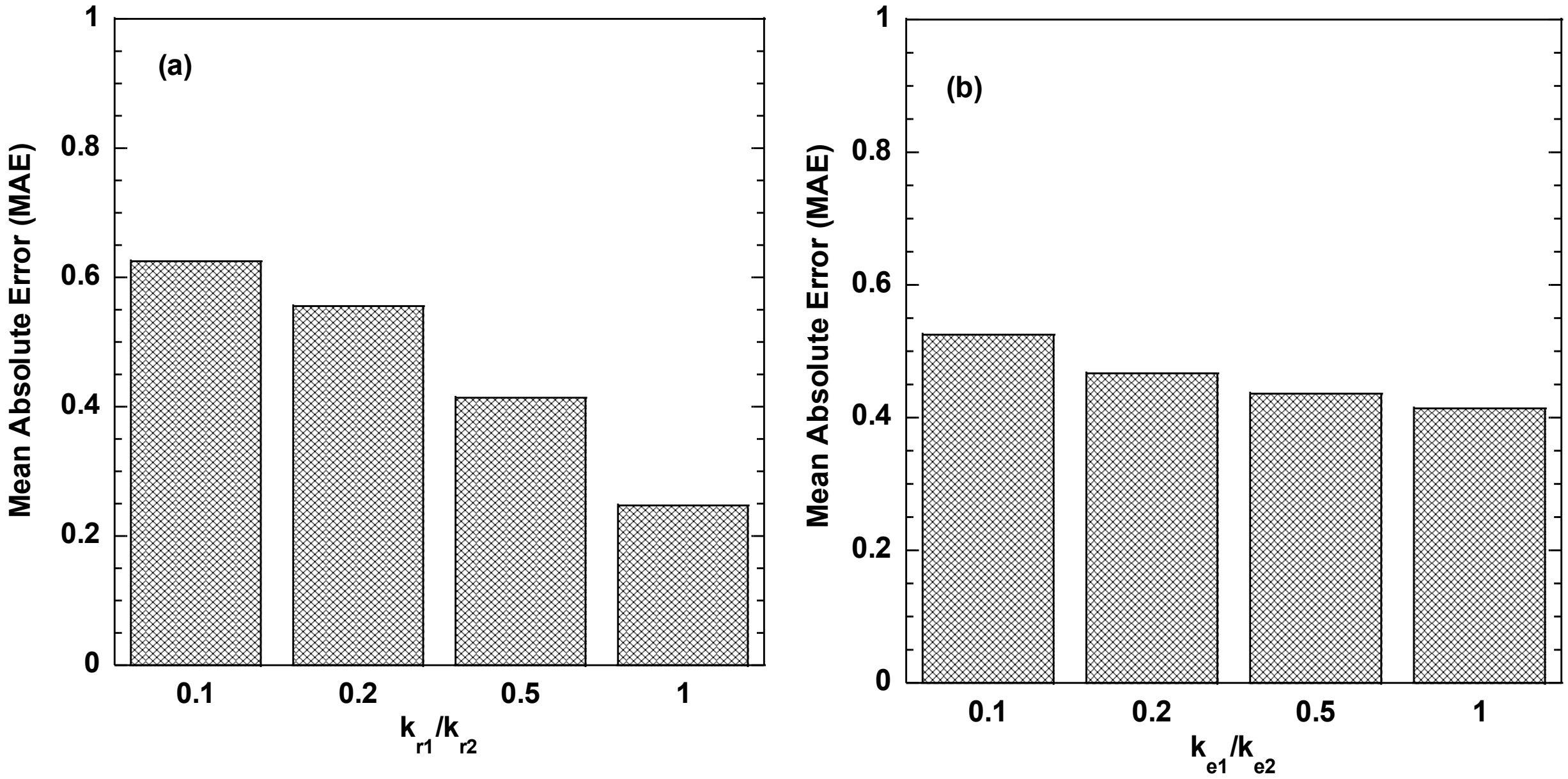

**Fig. 18.** Effect of conductivity ratios on the mean absolute error (MAE): (a) influence of $k_{r1}/k_{r2}$; (b) influence of $k_{e1}/k_{e2}$.

# 6. Summary and Conclusions

This study developed a dimensionless multi-domain physics-informed neural network framework for radial electro-osmotic consolidation considering smear effects under combined vacuum pressure and time-dependent surcharge loading. The smear and undisturbed zones were represented by two separate sub-networks and coupled through pressure and total radial flux continuity at the interface. Three formulations, including Std-PINN, Mod-PINN, and Mod-HC-PINN, were evaluated against FEM reference solutions under four loading cases with increasing complexity including constant vacuum (C1), exponential vacuum (C2), exponential vacuum with ramp surcharge (C3), and exponential vacuum with cyclic haversine surcharge (C4). The main conclusions are as follows:

1) The proposed multi-domain PINN framework effectively represents radial electro-osmotic consolidation with smear effects by assigning separate neural networks to the smear and undisturbed zones. The interface continuity constraints allow the model to capture the gradient transition caused by hydraulic permeability contrast between the two zones, avoiding the difficulty of using a single network to approximate a solution with discontinuous derivative behavior.

2) For the constant-vacuum case (C1), the modified gated architecture substantially improves the accuracy and training stability compared with the standard MLP architecture. The Std-PINN shows larger errors near the cathode and smear-zone interface, where steep radial pressure gradients occur, whereas the Mod-PINN reduces the MAE from 2.01 to 0.32 kPa. This demonstrates that the multiplicative gating mechanism enhances the model's ability to resolve localized high-gradient features in electro-osmotic consolidation.

3) For time-dependent vacuum and surcharge loading cases, the soft-constrained Mod-PINN becomes less effective than in constant-vacuum case. The gradient cosine-similarity analysis indicates that this limitation is not mainly caused by direct conflict between the boundary-condition and initial-condition losses, but rather by a multi-objective capacity burden under time-dependent boundary loading. By embedding the cathode boundary condition and initial condition directly into the output structure, the Mod-HC-PINN removes these two loss terms from the optimization objective and allows training to focus on the PDE residual, anode boundary condition, and interface continuity. As a result, the Mod-HC-PINN consistently outperformed the soft-constrained Mod-PINN under time-dependent vacuum and surcharge loading, reducing the MAE to 0.43, 0.41, and 0.27 kPa for Cases C2–C4, respectively.

4) Sensitivity analyses confirm that the proposed Mod-HC-PINN remains robust over practical ranges of network depth, width, collocation density, and hydraulic/electro-osmotic permeability contrasts. Accurate results were obtained using moderate network sizes and 30,000 collocation points, while the framework maintained low prediction errors under varying smear-zone permeability ratios.

The proposed PINN framework provides an accurate and stable mesh-free approach for modeling electro-osmotic radial consolidation under complex time-dependent loading histories. Future work may extend the proposed framework to more realistic electro-osmotic consolidation conditions by incorporating nonlinear soil behavior, coupled drainage mechanisms, and field-scale validation. The present study assumes constant hydraulic permeability, electro-osmotic permeability, and compressibility. Future developments should incorporate void-ratio-dependent permeability, compressibility, and electro-osmotic conductivity to better represent large-strain consolidation behavior in soft clays. Additional electrokinetic mechanisms, including voltage degradation, electrode resistance, electrochemical reactions, pH variation, and temperature effects, may also be incorporated to improve the physical realism of long-duration electro-osmotic treatment. The current radial-only formulation can also be extended to coupled radial–vertical drainage, multilayered soil profiles, and partially drained boundaries, which are more representative of field applications involving vertical drains or electric vertical drains. In addition, the proposed framework can be further developed for inverse analysis using laboratory or field monitoring data to identify uncertain consolidation parameters, loading histories, and smear-zone properties. Validation against laboratory and field-scale electro-osmotic vacuum preloading tests would further assess the generalization capability and practical reliability of the proposed Mod-HC-PINN framework under real ground-improvement conditions.

**Credit author statement**

Dong Li: Methodology, Formal analysis, Investigation, Writing the original draft, Reviewing and editing the manuscript; Yapeng Cao and Yang Lu: Reviewing and editing the manuscript, Funding acquisition, Methodology; Shuai Huang and Yujun Cui: Conceptualization, Project administration, Supervision, Software training, Validation; Haiping Fu and Wei He: Visualization, Reviewing and editing the manuscript.

**Competing interests statement**

The authors declare there are no competing interests.

**Data availability statement**

Data generated or analyzed during this study are available from the corresponding author upon reasonable request.

**Acknowledgements**

The authors gratefully acknowledge the support of National Natural Science Foundation of China (Grant No. 42401176), National Natural Science Foundation of China (Grant No. 52279099).